\documentstyle[pre,aps,epsf,twocolumn]{revtex}
\begin{document}
\def\bea{\begin{eqnarray}}
\def\eea{\end{eqnarray}}
\def\a{\alpha}
\def\D{\langle l \rangle}
\def\p{\partial} 
\draft
\title{The triangular Ising antiferromagnet in a staggered field}  
\author{Abhishek Dhar} 
\address{Condensed Matter Theory Unit\\ Jawaharlal Nehru Centre for
Advanced Scientific Research\\ Bangalore 560064, India}
\author{Pinaki Chaudhuri and Chandan Dasgupta~\cite{jnc}}
\address{ Department of Physics\\ Indian Institute of Science\\
Bangalore 560012, India.\\}
\date{\today}
\maketitle
\begin{abstract}
We study the equilibrium properties of the nearest-neighbor
Ising antiferromagnet on a triangular lattice in the presence of
a staggered field conjugate to one of the degenerate ground
states. Using a mapping of the ground states of the model
without the staggered field to dimer coverings on the dual
lattice, we classify the ground states into sectors specified by
the number of ``strings''. We show that the effect of the
staggered field is to generate long-range interactions between
strings.  In the limiting case of the antiferromagnetic coupling
constant $J$ becoming infinitely large, we prove the existence
of a phase transition in this system and obtain a finite lower
bound for the transition temperature. For finite $J$, we study
the equilibrium properties of the system using Monte Carlo
simulations with three different dynamics. We find
that in all the three cases, equilibration times for low field
values increase rapidly with system size at low temperatures.
Due to this difficulty in equilibrating sufficiently large
systems at low temperatures, our finite-size scaling analysis of
the numerical results does not permit a definite conclusion about 
the existence of a phase transition for finite values of $J$.
A surprising feature in the system is the fact that unlike 
usual glassy systems, a zero-temperature quench almost
always leads to the ground state, while a slow cooling does not.
\end{abstract}



\section{Introduction}

The triangular Ising antiferromagnet (TIAFM), described by the
Hamiltonian
\bea
H= J \sum_{ \langle i,j \rangle } s_i s_j   ~~~~~ J > 0,
\label{eqn1}
\eea
where $s_i = \pm 1$ and 
$\langle i,j \rangle$ denotes nearest-neighbor sites on a
triangular lattice, provides an interesting example of a
frustrated system without disorder. Unlike the nearest-neighbor
Ising antiferromagnet on a square lattice, this model does not
have a finite-temperature phase transition. It has an
exponentially large number of degenerate ground states, which
implies that the zero-temperature entropy per spin is finite.
The zero-field partition function can be computed exactly,
leading to the result $S(T=0)=0.3383...$ \cite{wann} for the
zero-temperature entropy per spin. At zero temperature, the
system is critical and the two-spin correlation function decays
as a power law, $c(r) \sim cos(2 \pi r/3)/r^{1/2}$, along the
three principal directions~\cite{stephen}. The ground states of
the TIAFM can be mapped exactly to dimer coverings on the dual
lattice which is hexagonal~\cite{nien}. Using this mapping, it is
possible to classify the ground states into sectors specified by
the number of ``strings'' which represent the difference between
two dimer coverings.

The exponential degeneracy of the ground state of the TIAFM can
be removed in various ways, e.g., by choosing different coupling
constants along the three principal directions, or by
introducing a uniform field. Both these cases have been
extensively studied.  For anisotropic couplings, the problem is
exactly solvable~\cite{hout} and one finds a usual Ising-like
second-order phase transition except in some special cases for
which the transition temperature goes to zero. In the case
of a uniform field, simulations and renormalization-group
arguments~\cite{kinz} indicate that there is a second-order
transition belonging to the $3$-state Potts model universality
class.

A particularly interesting special case is the limit in
which the system is restricted to remain within the manifold of
the TIAFM ground states. This can be achieved by making the
coupling constant $J$ infinitely large. One then considers the
effects of degeneracy breaking terms. In this limit, the nature
of the transition changes. In the case of anisotropic couplings,
the transition changes from Ising-like to Kastelyn-type 
($K$-type)~\cite{blote}. Below $T_c$, the system freezes into the
ground state and the specific heat vanishes identically. As
$T_c$ is approached from the high-temperature side, the specific
heat shows a $(T-T_c)^{-1/2}$ singularity. In the case of a
uniform applied field, the transition is believed to be of
Kosterlitz-Thouless type~\cite{nighting}. This case is treated
by first mapping the problem to a solid-on-solid model and then
using renormalization-group arguments.

In this paper, we study the behaviour of the TIAFM in the
presence of a staggered field chosen to be conjugate to
one of the ground states. Our work is motivated in part by
similar studies on glassy systems~\cite{silvio} with
exponentially large number of metastable states. These studies
consider the thermodynamic behavior of such systems in the
presence of a field conjugate to a typical configuration of an
identical replica of the system.  As the strength of the field
is increased from zero, the system is found to undergo a
first-order transition in which the overlap with the selected
configuration changes discontinuously. This transition is driven
by the competition between the energy associated with the field
term and the configurational entropy arising from the presence
of an exponentially large number of metastable states.  Like
these glassy systems, the TIAFM has frustration and an
exponentially large number of ground states.  Thus it is of
interest to investigate whether a similar behaviour is present
in the TIAFM which is a simpler model with no externally imposed
quenched disorder. Besides, the question of whether a phase
transition can occur in the TIAFM in the presence of an ordering
field is interesting by itself. For systems with a finite number
of ground states, such as the purely ferromagnetic Ising model
and the Ising antiferromagnet on a bipartite lattice, it can be
proved that no phase transition can occur in the presence of
ordering fields~\cite{yang}. However, no such general proof
exists for systems with an exponentially large number of ground
states, and the question of whether a competition between the
energy associated with the ordering field and the extensive
ground-state entropy can drive a phase transition in such
systems remains open.

The staggered field considered by us is conjugate to a
ground state with alternate rows of up and down spins. In the lattice gas 
picture of the Ising model, this corresponds to an applied potential 
which is periodic in the  direction transverse to the rows.    
In the presence of the field, there are a large number of low-lying
energy states and this suggests the
possibility of an interesting phase transition as the temperature is
varied. We consider the case where the coupling constant $J$ is 
finite, as well as the limit $J \to \infty$. In the latter limit, one
considers only the set of states which are ground states of the
TIAFM Hamiltonian of Eq.~(\ref{eqn1}). 
In this limit, we show that the problem of evaluating
the partition function reduces to calculating the largest eigenvalue
of a one-dimensional fermion Hamiltonian with long-range coulombic
interactions. We have not been able to solve this problem but
have obtained a finite lower bound for the
transition temperature. The transition appears to be $K$-type.

For finite $J$, we have studied the equilibrium behavior of the
system by Monte Carlo (MC) simulations using three different kinds of
dynamics: (1) single-spin-flip Metropolis dynamics, (2) cluster
dynamics and (3) ``string'' dynamics in which all the spins on a
line are allowed to flip simultaneously. We find that in all three cases,
equilibration times at low fields and low temperatures increase
rapidly with system size. The last dynamics is found to be the
most efficient one for equilibrating the system in this regime. 
Finite-size scaling analysis of the data for small fields
suggests the existence of a characteristic temperature near which the
correlation length becomes very large. However, because of the
long equilibration times, we
have not been able to study large enough systems to be able to
answer conclusively the question of whether this corresponds to 
a true phase transition.

One surprising finding of our study concerns zero-temperature
quenches of the system, starting from random initial
configurations. We show that the system almost always reaches the
ground state in such quenches. On the other hand, a slow cooling of the
system leads 
to a metastable state. This is contrary to what happens 
in usual glassy systems where a fast quench usually leads to the
system getting stuck in a higher energy state, while a slow cooling
leads to the ground state with a high probability

The paper is organized as follows. In section II, we consider
the TIAFM in zero field and describe the mapping from the ground
states to dimer coverings and the subsequent classification of
the ground states into sectors. Many of the results in this
section are well-known, but we have included them for the sake
of completeness.  Also our description is somewhat different
from the existing ones. In section III, we consider the TIAFM
with an applied staggered field in the limit $ J \to
\infty$. The mapping of this system to a one-dimensional fermion
model is described and a finite lower bound for the transition
temperature is derived. In section IV, we present our numerical
results for the equilibrium properties at finite $J$. These
results are obtained from exact numerical evaluation of averages
using transfer matrices and also through MC simulations. 
We also discuss the dynamic behaviour of the system under different MC
procedures. Section V contains a summary of our main results and a few
concluding remarks.

\section{ Mapping of TIAFM ground states to dimer coverings and
classification into string sectors}

The frustration of the TIAFM arises from the fact that it is impossible
to satisfy all three bonds of any elementary plaquette of the
triangular lattice. At most we can have two bonds satisfied. The
lowest energy configuration of the system is one in which
every elementary triangle is maximally satisfied. This condition can
be satisfied for a large number of configurations and for future
reference we shall denote the set of all such states by
${\cal G}$. We now show 
the correspondence between the ground states and dimer
coverings on the dual lattice. The dual 
lattice is formed by taking the centers of all the 
triangles. Consider any two triangles which share a bond. If the bond
is not satisfied, we place a dimer connecting the centers of the two
triangles. The fact that 
every triangle has one and only one unsatisfied bond implies
that every point of the dual lattice forms the end-point of one
and only one dimer. Hence we obtain a dimer covering. 
This mapping is
not unique, since flipping all spins in any given spin configuration
leads to the same dimer covering.
\vbox{
\vspace{0.5cm}
\epsfxsize=8.0cm
\epsfysize=6.0cm
\epsffile{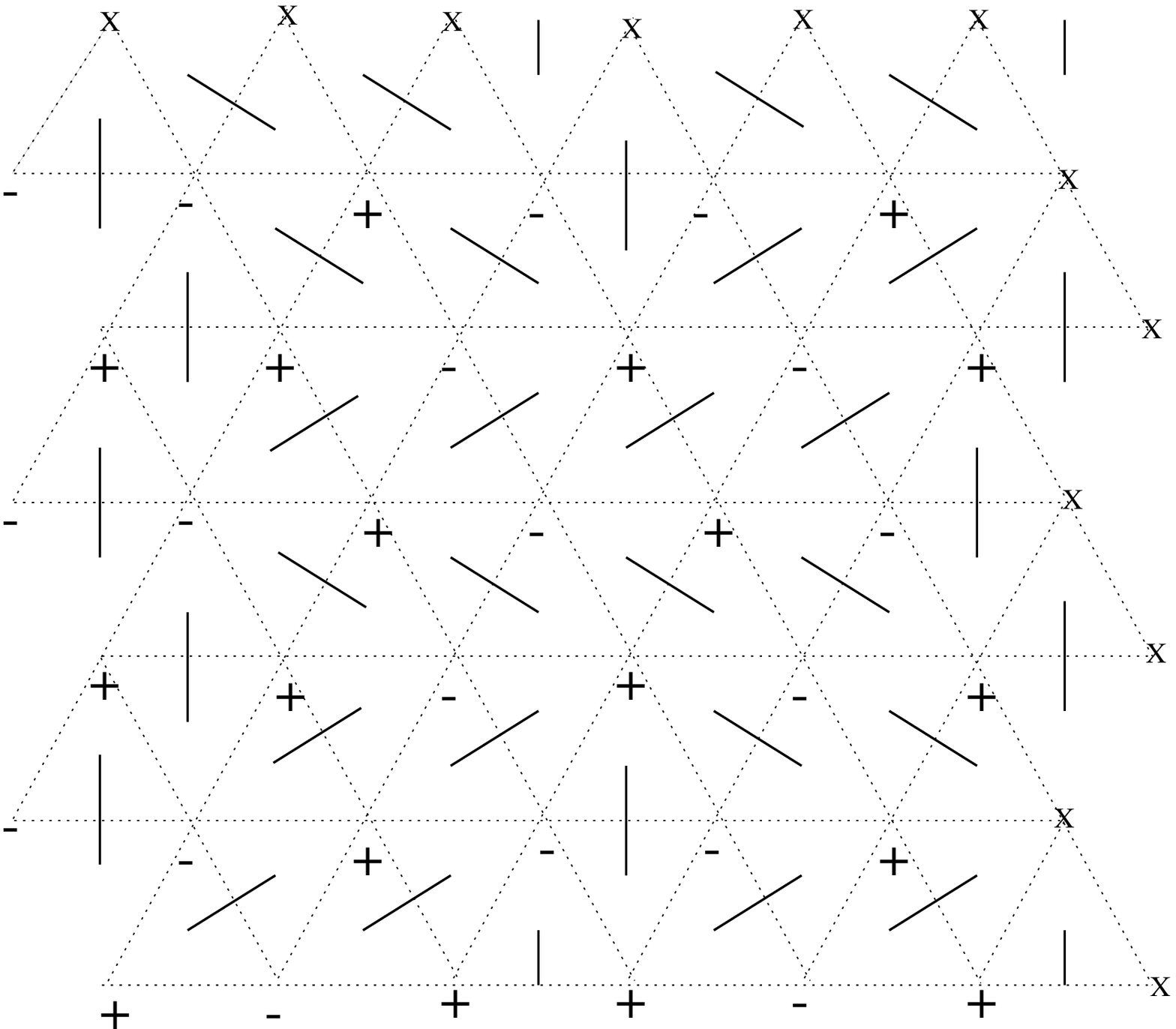}
\begin{figure}
\caption{\label{ground} A ground state configuration and the
corresponding dimer covering for a $6 \times 6$ lattice. Periodic
boundary conditions are applied in the horizontal and vertical
directions. The crosses correspond to repeated points. 
}  
\end{figure}}
In Fig.~\ref{ground} we show a
ground-state configuration and the corresponding dimer covering. 
Another dimer covering which corresponds to a
ground state with alternate rows of up and down spins is shown in
Fig.~\ref{stand}. We shall 
call this the standard configuration. It is important to choose the
boundary conditions in a convenient manner and we follow the convention
used in Fig.~\ref{ground} with periodicity in the $x$ and $y$ directions.

A useful classification of the ground states is obtained by
superposing the standard dimer configuration with any other dimer 
configuration.  This results in string configurations as shown, 
for example, in Fig.~\ref{string} which is obtained by superposing 
the standard configuration of Fig.~\ref{stand} with the configuration of
Fig.~\ref{ground}.  
Clearly there is a one-to-one correspondence between string and dimer
configurations. 

It is easy to prove the following points: 
(i) the number of strings
passing though every row is conserved; (ii) the strings do not
intersect; (iii) the number of strings can be any even number from $0$
to $L$, where $L$ is the number of spins in a row; (iv) the periodic
boundary conditions mean that the strings have to match at the
boundaries and form closed loops. 

We classify the ground states into different
sectors, with each sector specified by the number of strings. 
The number of states in each sector can be counted exactly using
transfer matrices. Let us label the bonds on successive rows of the
lattice in the manner shown in Fig.~\ref{transf}. The position of the strings
on each row is specified by the set of numbers $\{ b_1,b_2,...b_n\}$,
where $b_k$ gives the position of the $k$th string. Note that $\{ b_k \}$
give the positions of the satisfied bonds in a row. In a sector with
$n$ strings we consider the  ${}^L C_n \times {}^L C_n$ matrix which
has non-vanishing entries equal to one if  
the two states can be connected by string configurations.   
We need two different transfer matrices, namely $T^{(1)}$, 
which transfers from
odd numbered rows to even numbered ones and $T^{(2)}$, which transfers
from even to odd ones. The total number of states in any given sector
is then given by: 
\vbox{
\vspace{0.5cm}
\epsfxsize=8.0cm
\epsfysize=6.0cm
\epsffile{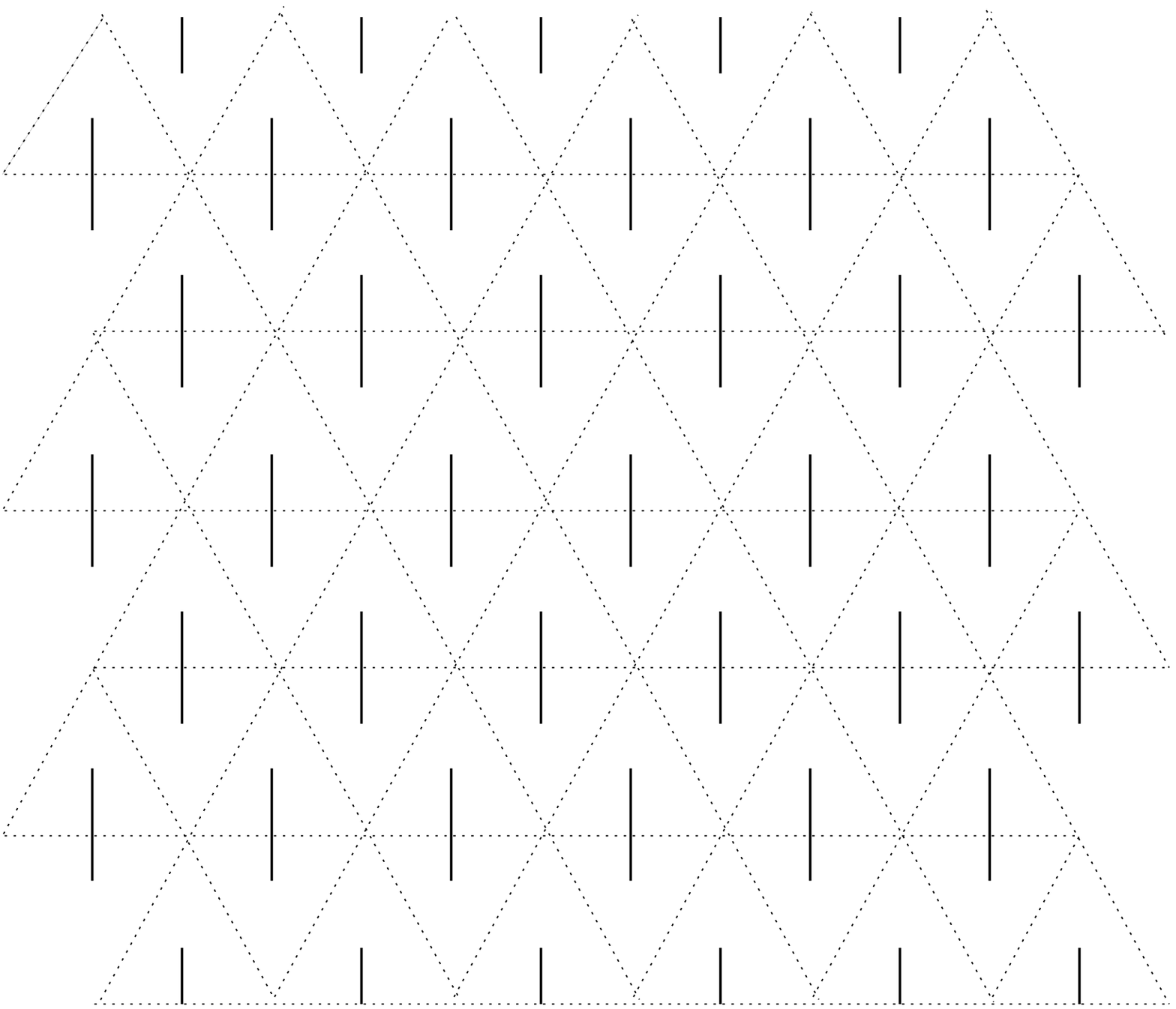}
\begin{figure}
\caption{\label{stand} The standard configuration of dimers.  
}  
\end{figure}}
\vbox{
\vspace{0.5cm}
\epsfxsize=8.0cm
\epsfysize=6.0cm
\epsffile{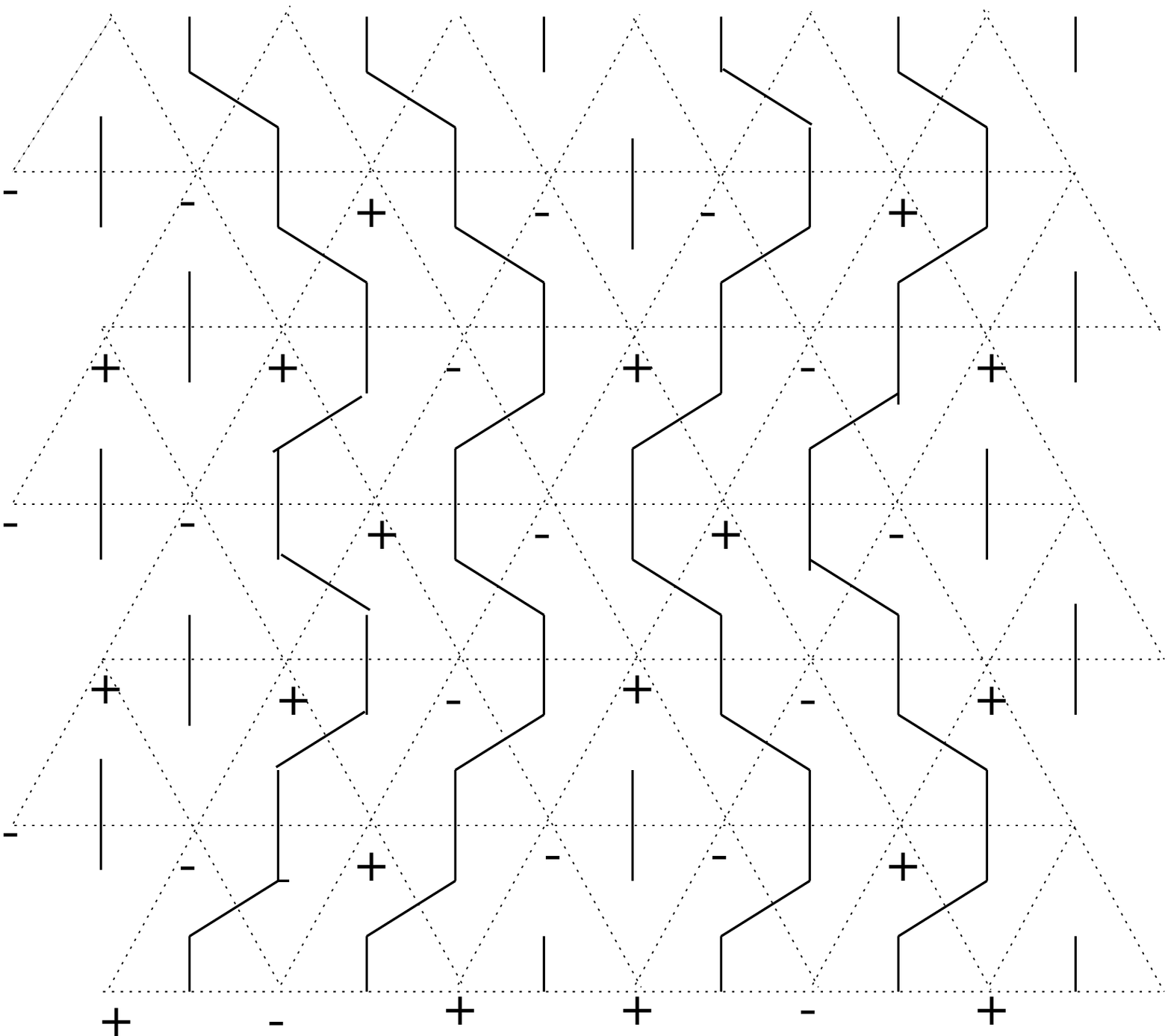}
\begin{figure}
\caption{\label{string} A configuration of strings obtained by
superposing the dimer configurations in Fig.~\ref{ground} and
Fig.~\ref{stand}. 
}  
\end{figure}}
\vbox{
\vspace{0.5cm}
\epsfxsize=8.0cm
\epsfysize=4.0cm
\epsffile{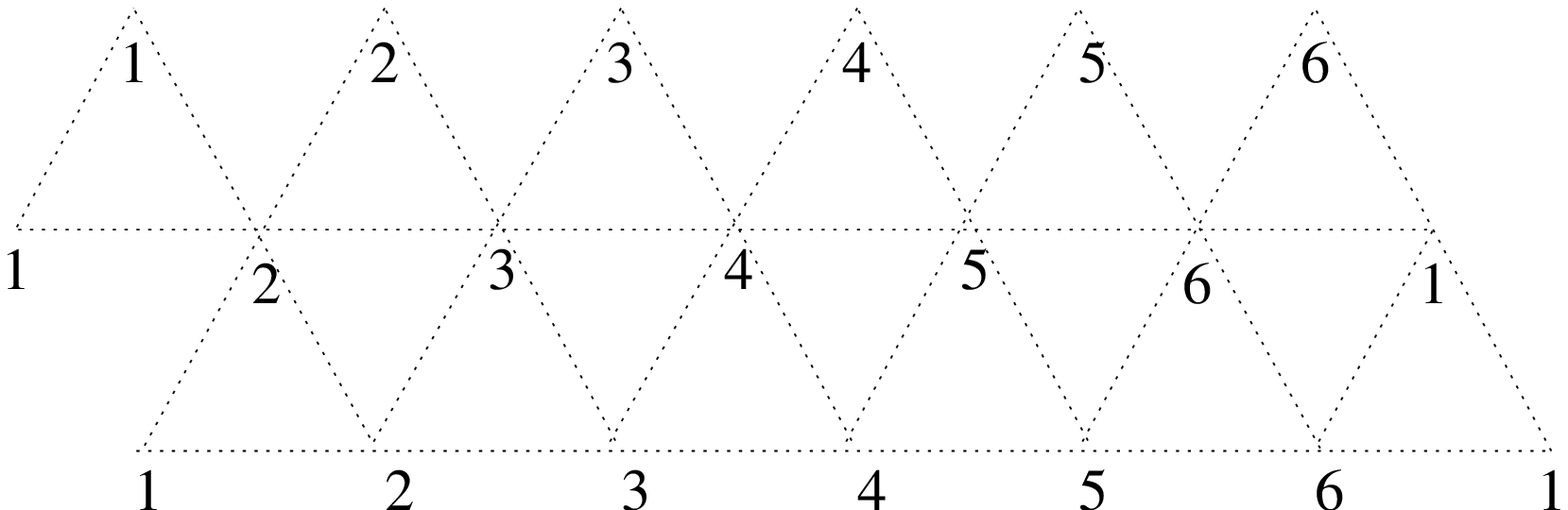}
\begin{figure}
\caption{\label{transf} Labelling of successive rows on a $6 \times 6$ 
lattice.
}  
\end{figure}}
\newpage
\bea
{\cal N}(n)=Tr (T^{(1)} T^{(2)})^{L/2},
\eea
where we choose, for convenience, the length of the lattice, $L$, to
be even. 

As an example let us consider the transfer matrix in the two string 
sector. This is given by 
\bea
T^{(1)}_{(l_1,l_2)\mid (l_3,l_4)} = \delta_{l_1,l_3} \delta_{l_2,l_4}  + 
\delta_{l_1,l_3-1} \delta_{l_2,l_4}  + \delta_{l_1,l_3}
\delta_{l_2,l_4-1} \nonumber \\
+ \delta_{l_1,l_3-1} \delta_{l_2,l_4-1} ~~~~~\rm{for}~~~l_2 \neq l_1+1,
\nonumber    \\
T^{(1)}_{(l_1, l_1+1) \mid (l_3, l_4)} = \delta_{l_1, l_3} \delta_{l_1+1,
l_4} +\delta_{l_1, l_3} \delta_{l_1+1, l_4-1} \nonumber \\ 
+ \delta_{l_1,l_3-1} \delta_{l_1+1, l_4-1}.
\eea
The matrix is diagonalized by the antisymmetrized plane-wave
eigenstates 
\bea
a_{l_1,l_2}= e^{i (q_1 l_1 + q_2 l_2)}-e^{i (q_1 l_2 + q_2 l_1) } ~,~
q_1 < q_2.
\eea
The periodic boundary condition leads to the following values for the
wave vectors: $q_i= (2 n_i+1) \pi /L$, with $n_i=0,1,2,...L-1$.
The eigenvalues are given by 
\bea
\lambda^{(1)}_{\bar q}=(1+e^{i q_1}) (1+e^{i q_2}).
\eea
The matrix $T^{(2)}$ has the same set of eigenvectors while the eigenvalues 
are given by
\bea
\lambda^{(2)}_{\bar q}=(1+e^{-i q_1}) (1+e^{-i q_2}).
\eea  

The results for the two-string sector can be generalized to any of the other
sectors. The transfer matrices $T^{(1)}$ and $T^{(2)}$ in any sector
are diagonalized by antisymmetrized plane wave states. This just
reflects the fact that the strings can be thought of as the world
lines of non-interacting fermions. The eigenvalues in the $n$-string
sector are:
\bea
\lambda^{(1)}_{\bar q} &=& \prod_{k=1}^{n} (1+e^{i q_k}),  \nonumber	\\
\lambda^{(2)}_{\bar q} &=& \prod_{k=1}^{n} (1+e^{-i q_k}), 
\eea
with $q_k$s as before.  
The number of states in the $n$-string sector is thus given by:
\bea
{\cal N}(n) &=& Tr (T^{(1)} T^{(2)})^{L/2} \nonumber \\
&=& \sum_{q_1 < q_2 ....q_n} [ \prod_{k=1}^{n} (1+e^{i q_k}) 
(1+e^{-i q_k}) ]^{L/2}
\eea
In the large $L$ limit, only the dominant term in the above sum
contributes and we finally obtain:
\bea
{\cal N}(p) &=& e^{L^2 \alpha(p)}, \nonumber \\
\alpha(p) &=& p \ln 2+\frac{2}{\pi} \int_0^{\pi p/2} dx \ln(\cos (x)) ,
\label{ssec}
\eea
where $p=n/L$ is the fraction of strings (``string density''). 
Thus every sector with non-zero $p$ has an exponentially large number 
of states.

We note that the function $\alpha (p)$ is peaked at $p=2/3$ and the
entropy of this sector, $S=\alpha (2/3)$, reproduces the well-known
result of Wannier for the zero-temperature entropy of the TIAFM. Thus
we have rederived Wannier's result and also shown that most of the states
are in the sector with  string density equal to $2/3$.

\section{ Splitting of levels in the presence of a staggered field: The $J
\to \infty$ limit}

In the presence of a staggered field $h$ that is conjugate to one of
the ground states of the TIAFM, the macroscopic degeneracy of the
ground state is lifted. The field we consider is conjugate to the 
state corresponding to the standard dimer configuration
(Fig. \ref{stand}). There are two
such spin configurations and we choose the one which has all up spins on
the first row. Note that in the presence of the field, any two states
related by the flipping of all the spins have the same string representation
but different energies. To remove this ambiguity, we use an additional
label for the string states, which we take as the sign of the first
spin in the first row. The spin configuration on any row is then fully
specified by the set ${(s, b_1, b_2,...b_n)}$. 

Let us now look at the effect of the field in splitting the energy
levels in each sector. In the zero-string sector there are two states,
one corresponding to the ground 
state and the other, obtained by flipping all spins, to 
the highest energy state. The lowest energy states in 
the two-string sector can be generated by starting with the ground-state 
spin configuration and flipping a line 
\vbox{
\vspace{0.5cm}
\epsfxsize=8cm
\epsfysize=6.0cm
\epsffile{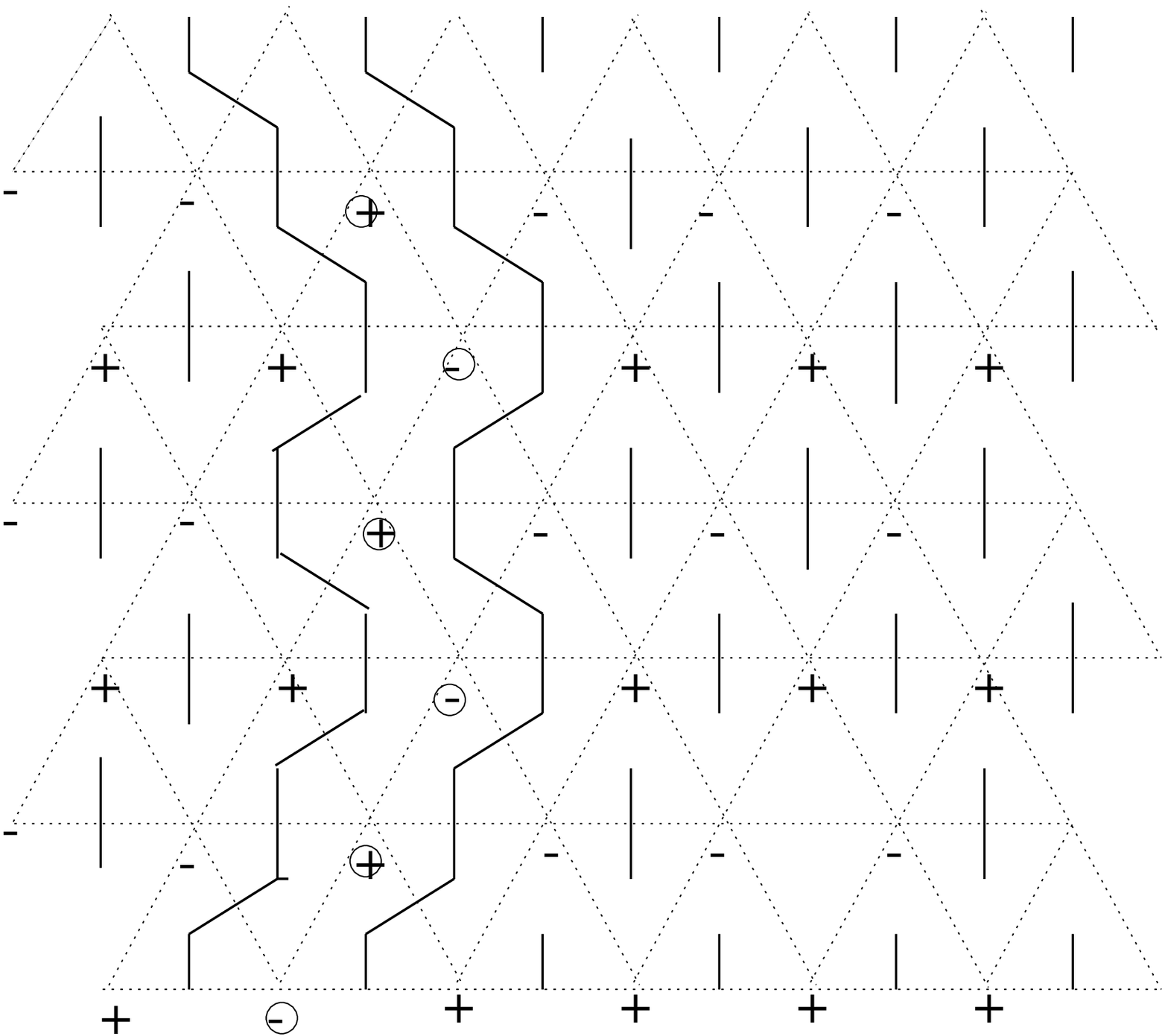}
\begin{figure}
\caption{\label{twostr} A configuration of two strings which
corresponds to the lowest energy state in this sector. 
This configuration is obtained by starting with the ground state and
flipping a line of spins (the circled ones).
The strings are
closely packed and all the spins in the region between them point opposite
to the local applied fields. 
}  
\end{figure}}
\vbox{
\vspace{0.5cm}
\epsfxsize=8.0cm
\epsfysize=5.5cm
\epsffile{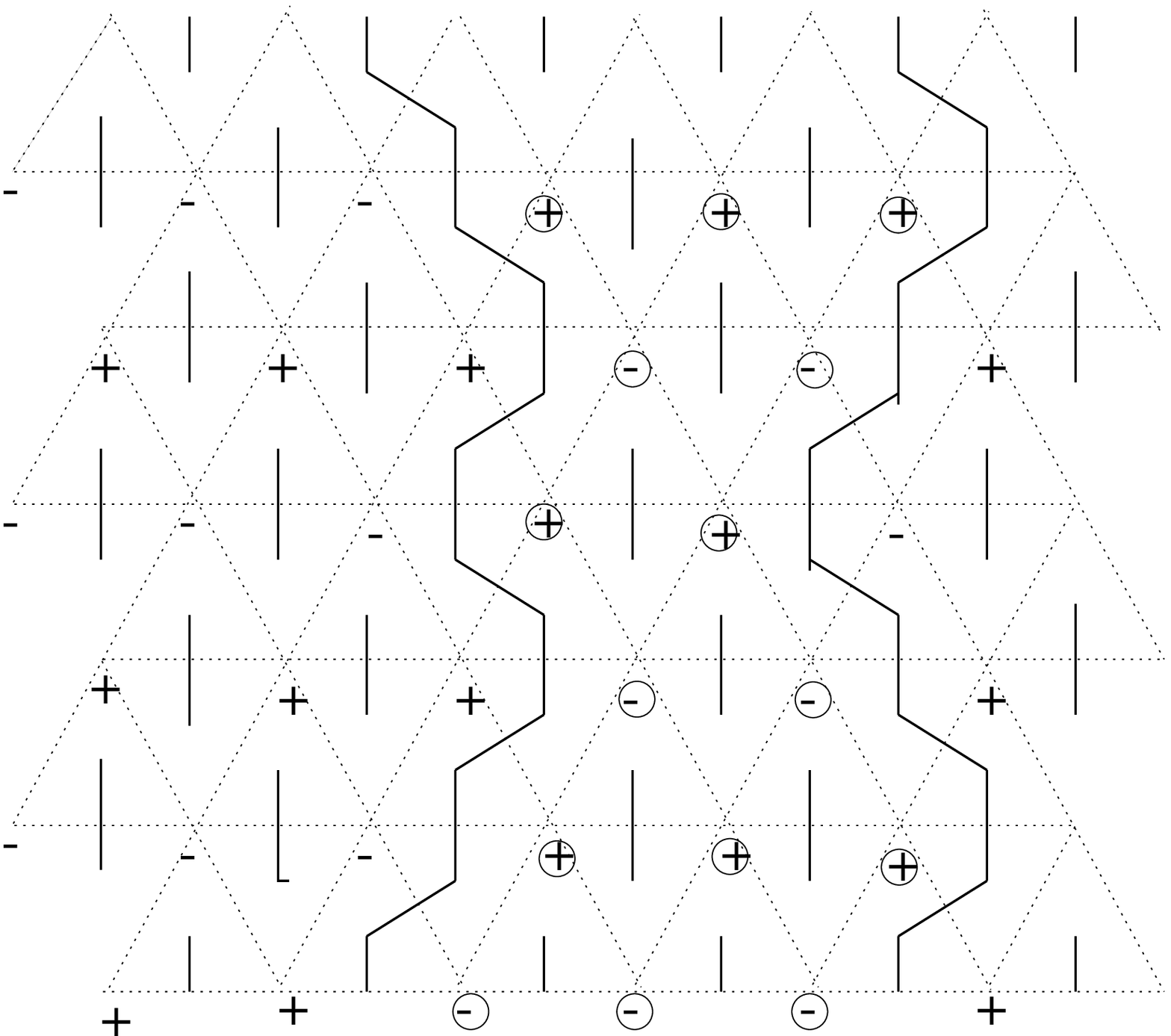}
\begin{figure}
\caption{\label{twostr2} A higher energy configuration in the
two-string sector. It can be seen that the strings divide the lattice into
two domains with the spins in one domain being along the applied field and
opposite to it (the circled spins) in the other domain.
}  
\end{figure}}
of spins as shown in Fig.~\ref{twostr}. Fig.~\ref{twostr2} 
shows a higher energy two-string state. Note that
the strings separate the lattice into two domains, one in which
all the spins
point along the staggered field directions and another in which they
point opposite to
the field. This is in general true for any $n$-string state where the
strings divide the lattice into $n$ domains, with spins in alternate
domains pointing along and opposite to the staggered fields.
The lowest energy configuration in any sector is clearly the state with 
alternate pairs of strings tightly packed.
For the sector in which the string density is $p$, the lowest 
energy per spin is 
\bea
e_g(p)=-(1-p) h,
\label{esec}
\eea
where for the case $J \to \infty$ being considered here, we have
subtracted the infinite constant energy term $-J$.   

Because of the conservation of the number of strings across rows, the
transfer matrix is block diagonal,
each block corresponding to a fixed string sector. In the zero field
case the strings are noninteracting and the problem reduced 
essentially to that of free fermions on a line. In the present case,
however, the energy increases when the separation between two strings is
increased. In fact it is easy to see that this case reduces to a
one-dimensional fermion problem in which
every alternate pair of fermions interact with each other via an
attractive linear potential. It is then no longer simple to diagonalize the
transfer matrix. 
However, through the following argument we prove the existence of a phase
transition and obtain a lower bound for the transition temperature.
At zero temperature, the system will be in the ground state in
the zero-string sector. As the temperature is increased, the entropic
factor associated with the other sectors becomes important and can cause
either a gradual or a sharp transition to other sectors. To
determine which of the two possibilities actually occurs, we consider the
simpler case where the strings 
do not interact and all configurations belonging to the sector 
with string density $p$ have the same
energy $N e_g(p)$ where $N = L^2$ is the total number of spins. 
Since  all the states in
this sector  have energies greater
than or equal to $N e_g(p)$ in the
interacting model, a sharp transition in the non-interacting case
implies a sharp transition in the interacting model. In particular, if the
non-interacting model exhibits a transition at temperature $T_c$, so
that it is frozen in the ground state in the zero-string sector for $T \le
T_c$, then the interacting model must also be in the ground state for
all $T \le T_c$. In other words, the transition temperature of the
non-interacting model provides a lower bound to the transition
temperature of the interacting model.

The partition function of the non-interacting model may be written as
\bea
Z &=& \sum_p e^{N \alpha(p)-\beta N e_g(p)} \nonumber \\
  &=&   e^{N (\alpha(p_m)-\beta e_g(p_m))}, 
\label{freeen}
\eea
where $\beta = 1/T$ and $p_m$ is the value of $p$ 
corresponding to the minimum of the function $f(p)=-\alpha(p)+\beta e_g(p)$. 
Using Eq.~(\ref{ssec}) and Eq.~(\ref{esec}), we get
\bea
p_m &=& 0, ~~~~~ T < T_c, \nonumber   \\
p_m &=& \frac{2}{\pi} \cos^{-1} (\frac{e^{h/T}}{2}), ~~~~T>T_c,
\label{pmeq}
\eea
with $T_c=h/\ln (2)$. Thus, there exists a sharp transition at a finite
temperature $T_c$, the number of strings being identically zero below this
temperature. In Fig.~\ref{free}, we 
show the dimensionless free energy function $f(p)$ at two different 
temperatures, one
above and one below $T_c$. It can be seen that for $T < T_c$, the
function $f(p)$ has its lowest value at $p=0$. The minimum of $f(p)$
moves 
\vbox{
\vspace{0.5cm}
\epsfxsize=8.0cm
\epsfysize=6.0cm
\epsffile{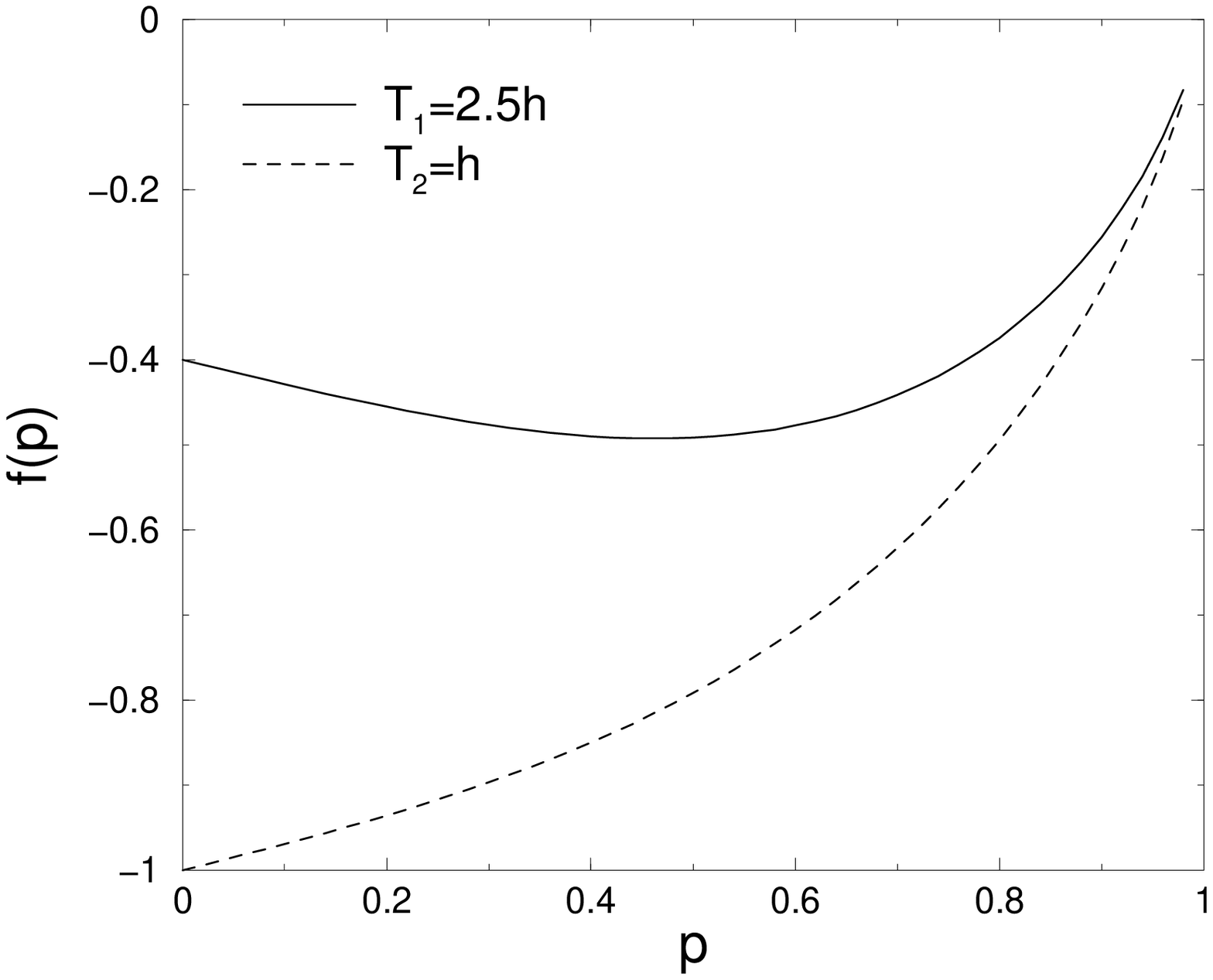}
\begin{figure}
\caption{\label{free} The dimensionless free energy $f(p)$ of the
non-interacting model, 
plotted as a function of the string density  $p$ at two different 
temperatures, $T_1 =2.5 h$ which is above $T_c$, and $T_2 = h$, which
is below $T_c$.
}  
\end{figure}}
continuously away from $p=0$ as the temperature is increased
above $T_c$, approaching $p = 2/3$ in the $T \to \infty$ limit. 

In Fig.~\ref{approx}, we have plotted $p_m$, the equilibrium value of
the string density obtained from Eq.~(\ref{pmeq}), as a
function of $T/h$. It is easy to see from Eq.~(\ref{pmeq}) that $p_m$
grows as $(T-T_c)^{1/2}$ as $T$ is increased above $T_c$. Since the
internal energy is proportional to $p_m$ in the non-interacting
model, the specific heat vanishes 
identically for $T < T_c$ and diverges as $(T-T_c)^{-1/2}$ for $T$
approaching $T_c$ from above. Thus we get a $K$-type transition which
is expected because of the equivalence of our system to dimer models. 
While this proves the existence of a transition in the interacting 
model too, it is not clear whether the nature of the transition is
the same. It is quite possible that the long-range interactions
\vbox{
\vspace{0.5cm}
\epsfxsize=8.0cm
\epsfysize=6.0cm
\epsffile{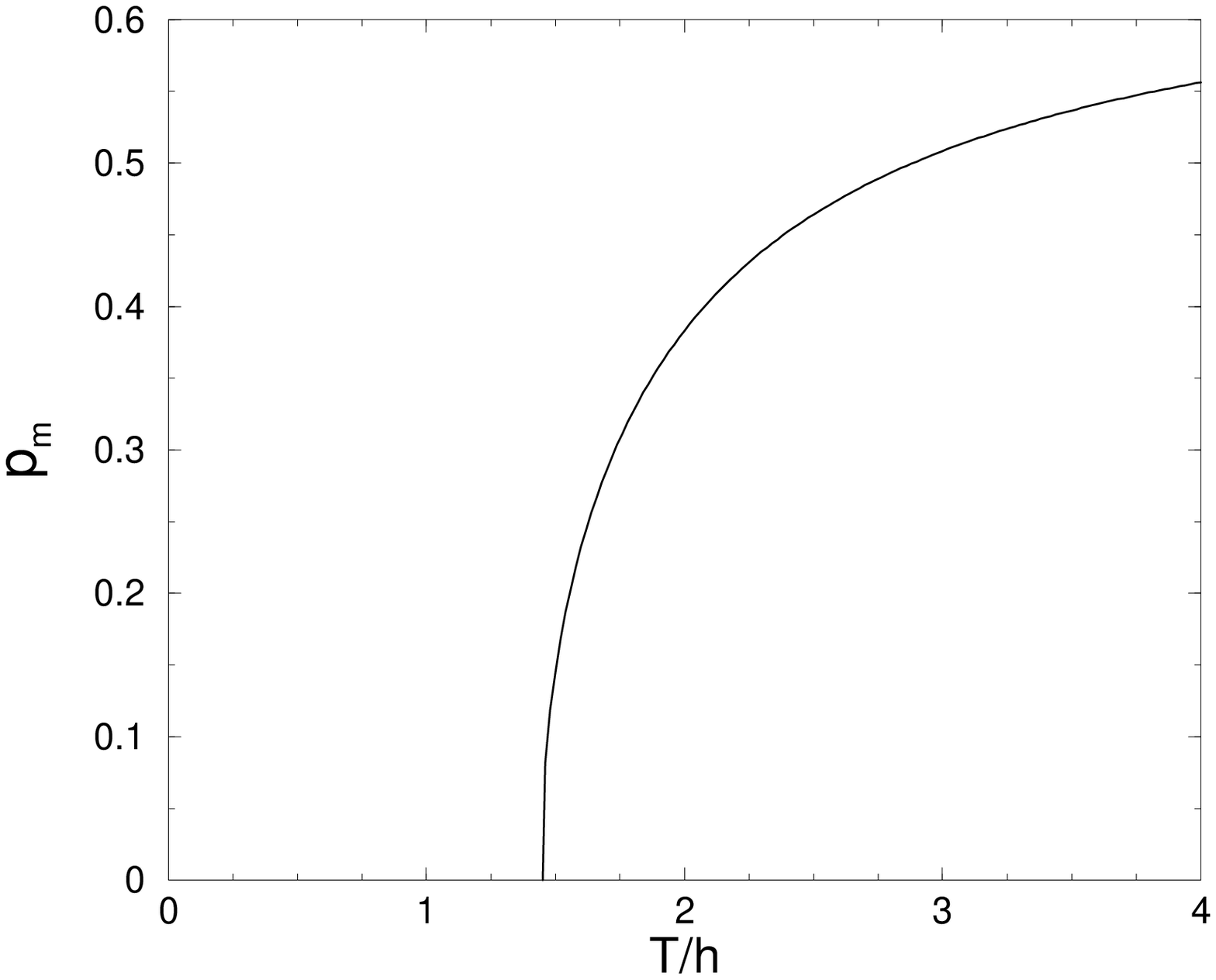}
\begin{figure}
\caption{\label{approx} The equilibrium string density $p_m$ plotted as
a function of the temperature $T$ (measured in units of $h$) for the 
non-interacting string model.
}  
\end{figure}}
between the strings would result in a transition in a different 
universality class. This issue is addressed in the next section.

It is interesting to compare our model with the model with 
anisotropic couplings studied by Bl\"{o}te and
Hilhorst~\cite{blote}. Consider the case when the
horizontal couplings have strength $(J-\Delta)$ and the remaining two
are of strength $J$. In the limit $J \to \infty$, we need to
consider only the states within ${\cal G}$. In this case too, the ground
state lies in the zero-string sector but is two-fold degenerate since
the up-down symmetry is retained. The excitations are again in the
form of strings but are non-interacting and so equivalent to the
excitations in the
simplified model considered by us. In fact the expression for the free
energy in Eq.~(\ref{freeen}) follows directly from Eq.~(2) in 
Ref.~\cite{blote} if we make the identification $h=2 \Delta$.

\section{MC simulations and transfer matrix calculations for finite $J$} 

For finite $J$, we have carried out MC simulations to
determine whether the phase transition persists and its nature
if it does. A problem with the
simulations is that equilibration times are very long for small
values of $h/J$ and $T/J$. We have tried to overcome this
problem by performing simulations with three kinds
of dynamics. However, even with the fastest dynamics, we have
been able to obtain reliable data only for relatively small system sizes
$(L \le 18)$.  We have also carried out exact numerical
evaluation of averages using transfer matrices for small
samples. The results obtained from these numerical calculations
are described below.

\subsection{Single-spin-flip Metropolis dynamics}

In Fig.~\ref{sflp} we show the results of a MC simulation using 
the standard single-spin-flip Metropolis dynamics~\cite{mc}. We
have plotted the staggered magnetization $m$ as a function of 
temperature $T$ for a heating run and a cooling run on a $6
\times 6$ system. The staggered field
and the coupling constant are set to $h=0.05$ and 
$J=1.0$, respectively (Unless otherwise stated, all the
numerical results reported in this section are for $J = 1.0$).
The data shown were obtained by averaging over $10^6$
MC steps per spin (MCS). The heating run was started from the
ground state in the zero-string sector and the cooling run
started from a random spin configuration. It is clear from the
data that even for this small system, equilibration is not
obtained for temperatures lower than about 0.3. We also examined
the states obtained by starting the system 
\vbox{
\vspace{0.5cm}
\epsfxsize=8.0cm
\epsfysize=5.7cm
\epsffile{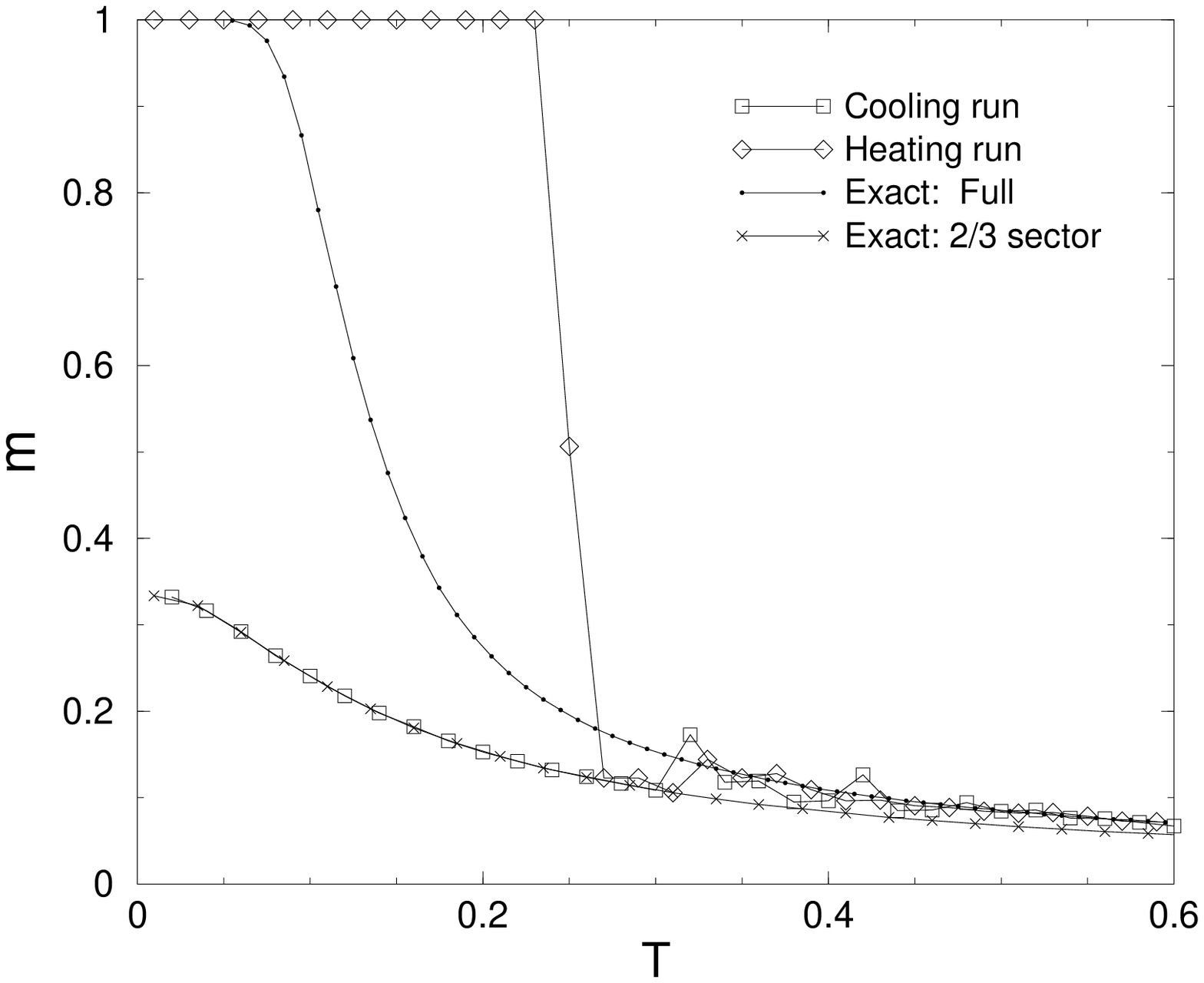}
\begin{figure}
\caption{\label{sflp} Results for the staggered magnetization
$m$, obtained from single-spin-flip MC heating
and cooling runs for a $6 \times 6$ system with $h=0.05$ and
$J=1$. Also shown are the results of exact numerical
evaluation of the staggered magnetization for $J=1$, and the
staggered magnetization in the $p=2/3$ sector for $J
\to \infty$.  
}  
\end{figure}}
in a random configuration
and then quenching it instantaneously to zero temperature. 
We find that the system then goes to the lowest-energy state in
one of the many sectors. For example, in the simulation corresponding
to Fig.\ref{sflp}, the system reached the zero-string ground
state. On heating, the system continues to be in the
zero-string sector until at some temperature value it jumps to the
high-temperature phase.
On the other hand, a slow cooling from the     
high temperature phase leads to the lowest energy state in the $p=(2/3)$
sector and the true zero-string ground state is not reached.

These results can be understood as follows. As discussed in
the preceding section, the ground state lies in the zero-string
sector, and the excitations within ${\cal G}$ from the ground state
correspond to the formation of an even number of strings.
The single-spin-flip dynamics is reasonably efficient in
exploring the states within a sector
with a fixed number of strings. However, at low temperatures, it
is extremely ineffective in changing the number of strings. In fact,
even with zero external field, the single-spin-flip dynamics at
zero temperature is non-ergodic and only samples 
states within a given sector. At finite temperatures, the only way to
change the number of strings is through moves which take the system
out of ${\cal G}$. These moves cost energy of order $J$. At low temperatures,
the probability of acceptance of such moves becomes extremely small.  
Thus in Fig.~\ref{sflp}, during the heating run, the system starts from the
ground state in the zero-string sector and stays stuck in it till the
temperature is sufficiently high. At high temperatures, the $p=2/3$
sector is most probable (note that at very high temperatures,
the string picture is no longer valid) and during the cooling run, the
system starts from this sector and again stays stuck in
this sector since the dynamics cannot reduce the number of
strings. Thus the cooling curve basically shows equilibrium properties
within the $p=2/3$ sector. 

We have verified the above picture by an exact
numerical evaluation of the staggered magnetization for a $6 \times 6$
system. This is done by numerically computing the two sums that
occur in the expression 
\bea
m = \frac{1}{N} \langle M \rangle= \frac{1}{N} \frac 
{Tr [M (V^{(1)} V^{(2)})^{L/2}]} {Tr [(V^{(1)}
V^{(2)})^{L/2}]},
\eea  
where $V^{(1),(2)}$ are the usual row-to-row transfer matrices and $M$ is a
diagonal matrix corresponding to the staggered 
magnetization. Similarly one can compute the staggered
susceptibility $\chi$ defined as
\bea
\chi = \frac{1}{N} [\langle M^2 \rangle - \langle M \rangle^2]. 
\eea
This exact evaluation can, however, be done only for small
systems since this
procedure involves using very large matrices. For finite $J$,
we have been able to do this calculation only for $L \le 6$.  
For $J \to \infty$, the transfer
matrices become block diagonal, and this means that one can perform
separately the computations in each block which are of smaller
size. In this case, we have been able to go up to system size $L=12$.
Note that in this limit, we can also compute the thermodynamic
properties in each sector. In Fig.~\ref{sflp}, we have plotted the exact
results for $m$ obtained from the full partition function with $J=1$,
as well as the results for $m$ in the $p=2/3$ sector for infinite $J$. 
It is readily seen that our
picture of the system getting stuck in the $p=2/3$ sector during the
cooling run is correct. 

The counter-intuitive results of the quenching process can also be
understood using the above picture.
After the quench, domains of spins pointing in and opposite to the
direction of the staggered field begin forming. Only spins on the
boundaries of the 
domains can flip, leading to motion of the domain
walls. This motion is biased, favouring the growth of the 
domains aligned with the staggered field. Now we recall that any
non-zero string configuration will have domains of 
misaligned spins spanning the entire lattice. Clearly it is extremely
unlikely that the biased domain growth process will lead to such
configurations. We have checked in our simulations that as the system
size is increased, the probability of the quench leading to the
zero-string sector approaches unity. To further clarify this process,
we show in Fig. \ref{quench1} different stages in the evolution of a
$24 \times 24$ system 
following a zero-temperature quench from a random initial
state. The field is set at the value $h=0.05$. It can be seen that the 
domains of misaligned spins rapidly vanish. On the other hand, in
Fig. \ref{quench2} we show a $T=0.4$ equilibrium spin configuration 
and the result of quenching it to $T=0$. In this case the system gets
stuck in the $p=2/3$ sector.
\vbox{
\vspace{0.5cm}
\epsfxsize=8.6cm
\epsfysize=4.3cm
\epsffile{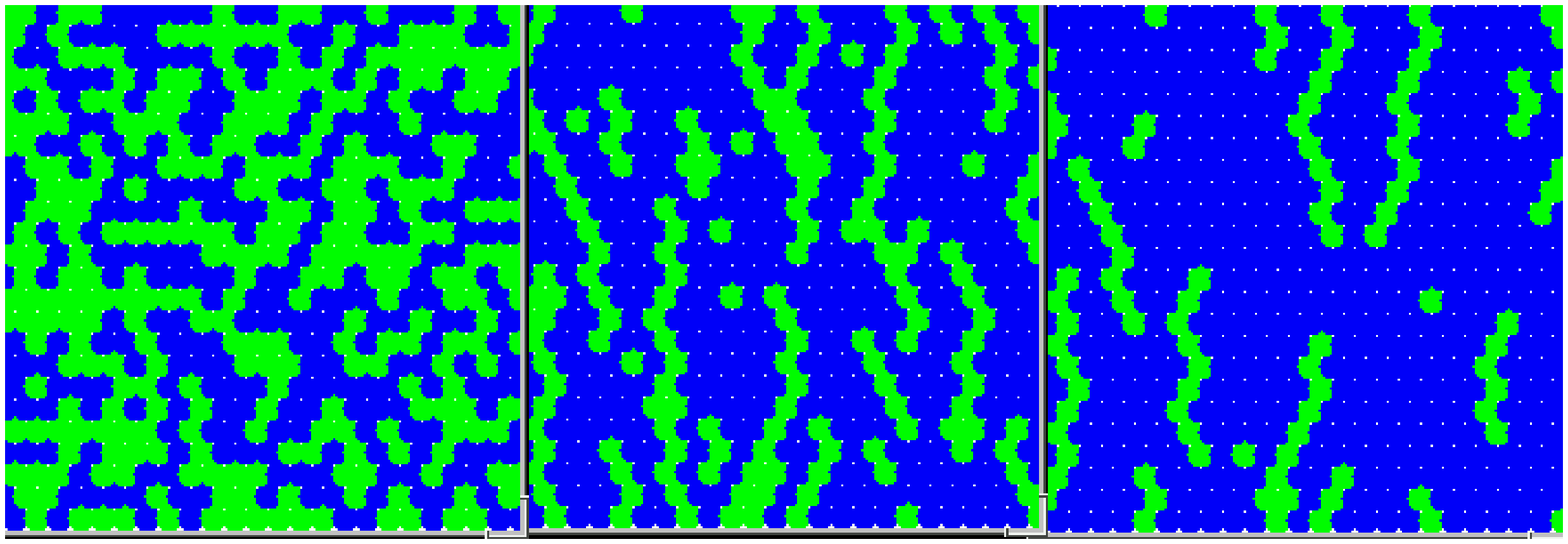}
\begin{figure}
\caption{\label{quench1} Three stages in the evolution of the system,
following a zero-temperature quench from a random initial state.
The first is the initial configuration and the other two are
configurations obtained after $2$ and
$4$ MC sweeps. The dark and bright regions indicate spins
pointing along and opposite to the direction of the staggered fields,
respectively.  
}  
\end{figure}}
\subsection{String dynamics}
To speed up the dynamics, it is necessary to be able to efficiently
change the number of strings. A straightforward way of doing this is
to introduce moves which 
\vbox{
\vspace{0.5cm}
\epsfxsize=8.0cm
\epsfysize=4.3cm
\epsffile{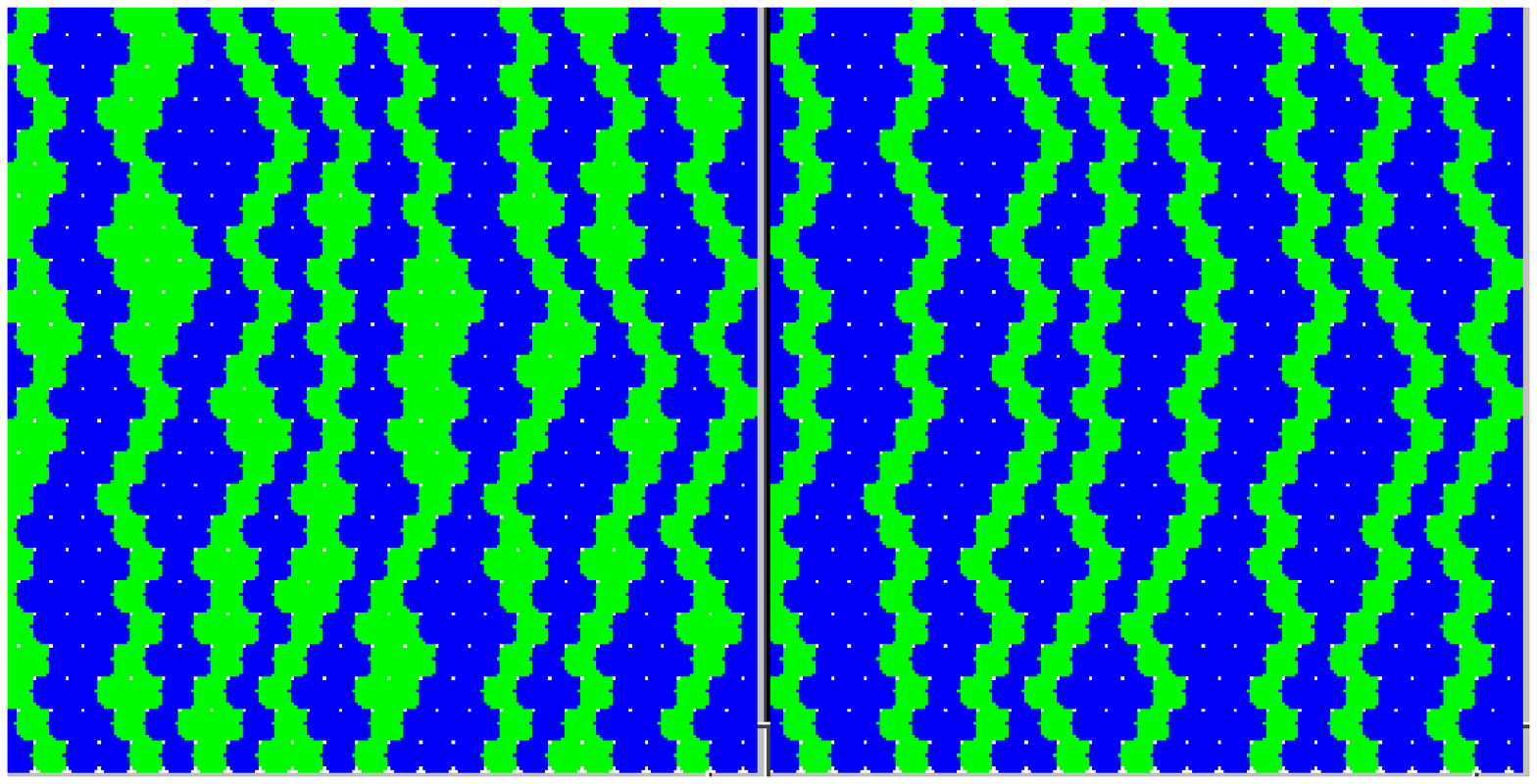}
\begin{figure}
\caption{\label{quench2} 
A configuration which is at equilibrium at $T=0.4$ and the
configuration resulting from quenching it to $T=0$. It can be seen that the
final configuration consists of tightly bound strings and is the
lowest energy state in the $p=2/3$ sector.  
}  
\end{figure}}
attempt to flip an entire vertical line of
spins. Such moves are accepted or rejected according to the usual 
Metropolis rules. Combining these moves with 
the single-spin-flip ones makes the dynamics ergodic at 
zero-temperature in the absence of the field. In Fig.~\ref{strflp},
we show the 
results of simulations with the string dynamics, again for a $6 \times 6$ 
system. The values of $J$ and $h$ are the
same as those for the data shown in Fig.~\ref{sflp}, and the
averaging is over the same number of MCS. 
The excellent agreement with the exact results
shows that equilibration times have been greatly reduced. We have also
shown in Fig.~\ref{strflp} simulation results for a $12 \times 12$
system. Again there is very good  
agreement with the exact results, which, as noted above, were obtained
by setting $J \to \infty$.     

To determine the existence of a phase transition, 
\vbox{
\vspace{0.1cm}
\epsfxsize=8.0cm
\epsfysize=6.4cm
\epsffile{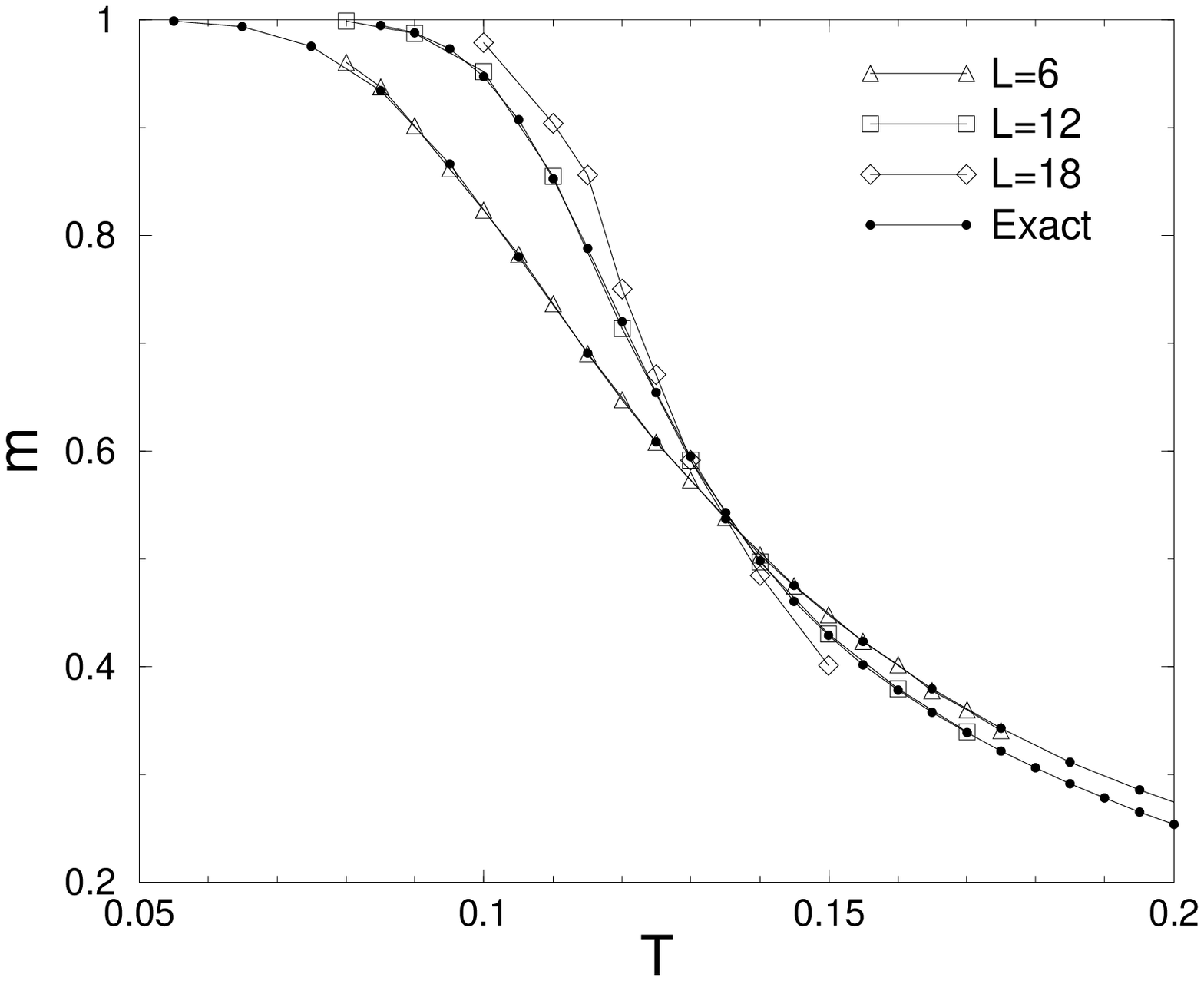}
\begin{figure}
\caption{\label{strflp} Staggered magnetization $m$ versus
temperature $T$ for $h=0.05$, $J=1$. The data for
system sizes $L=6$, $12$ and $18$ were obtained from
MC simulations using string dynamics. Exact transfer-matrix results for $L=6$
and for $L=12$ ($J \to \infty$) are also shown. 
}  
\end{figure}}
\newpage
\noindent we have performed simulations with the above dynamics and studied 
the dependence of the staggered susceptibility $\chi$ on the
system size for different values 
of the field. The results are summarized in
Figs.~\ref{h1}, \ref{h2} and \ref{h3}. 
The data in Fig.~\ref{h1} correspond to a low field value,
$h=0.05$. The number of MCS used for computing the
averages is $10^6$, $10^7$ and $4 \times 10^8$ for
the three system sizes, $L=6$, $12$ and $18$, respectively. 
For system sizes $L=6$ and $L=12$, we also show the exact
transfer-matrix results.
Even though the $L=12$ transfer matrix results are for
$J \to \infty$, we find very good agreement with the simulation
data. This is because excitations out of ${\cal G}$, which
involve energies of order $J$, are very much suppressed at the
low temperatures considered.
The $L=18$ MC data are not as smooth as the data for smaller sample
sizes, indicating that the errors in the calculation of averages
are significant in spite of averaging over a very large number of MCS.
Thus, even with the string dynamics, we have not been able to
attain equilibration for systems with $L > 18$. 
\vbox{
\vspace{0.5cm}
\epsfxsize=8.0cm
\epsfysize=7.0cm
\epsffile{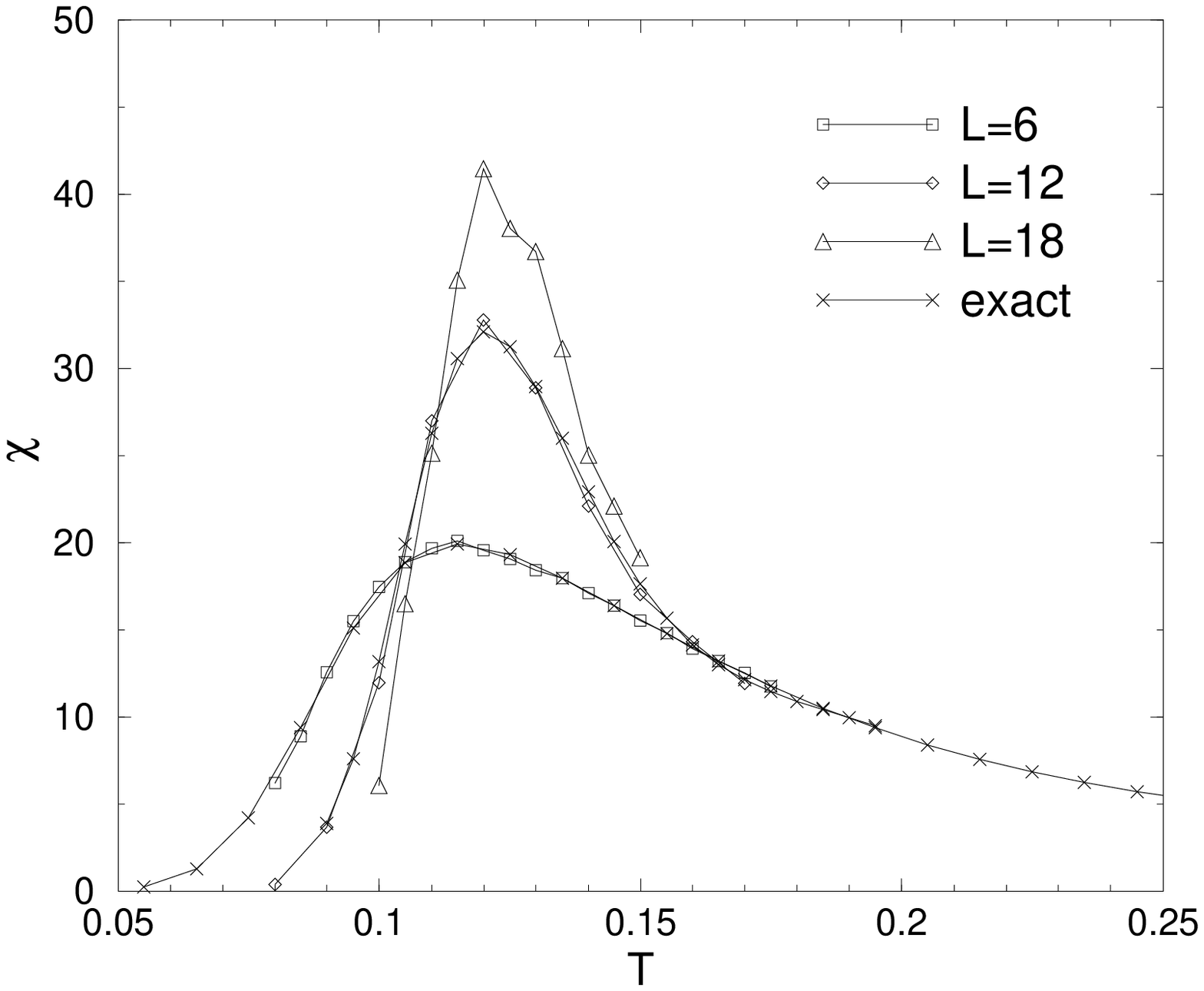}
\begin{figure}
\caption{\label{h1} Staggered susceptibility $\chi$ versus
temperature $T$ for $h=0.05$, $J=1$. The data for system sizes
$L=6$, $12$ and $18$ were obtained from
MC simulations using string dynamics. Exact transfer-matrix results for
$L=6$ and for $L=12$ ($J \to \infty$) are also shown. 
}  
\end{figure}}

The close agreement between the MC results for $J=1$ and the exact
transfer-matrix results for $J \to \infty$ indicates that the MC
results for the system sizes considered are representative of
the $J \to \infty$ limit. In section III, we established the
existence of a finite-temperature phase transition in this
limit. Our MC results indicate that this transition occurs near
$T \simeq 2.5 h$, which is substantially higher than the lower
bound, $h/\ln(2)$, derived in section III. To determine whether
this transition is K-type, we have examined the dependence of $\chi_p$,
the peak value of the staggered susceptibility $\chi$, on the
system size $L$. In the $J \to
\infty$ limit, the staggered susceptibility is proportional to
the specific heat which diverges as $(T-T_c)^{-1/2}$ in a K-type
transition. This implies that the susceptibility exponent
$\gamma=1/2$, and the correlation length exponent $\nu$ is equal
to $3/4$. According to standard finite-size scaling~\cite{fsc}, $\chi_p$
then should be proportional to $L^{\gamma/\nu} = L^{2/3}$. 
As shown in Fig.~\ref{max}, our numerical data are in
good agreement with this expectation. We, therefore, conclude
that our model undergoes a K-type transition in the $J \to \infty$ limit.

In Fig.~\ref{h2}, we show simulation results for an intermediate field
value, $h=0.25$. In this case, for system sizes $L=6$, $12$, $18$ and
$24$, equilibrium values were obtained by averaging over $2 \times
10^6$, $5 \times 10^6$, $2 \times 10^7$ and $5 \times 10^7$ 
MCS, respectively. As in the $h=0.05$ case, the 
\vbox{
\vspace{0.5cm}
\epsfxsize=8.0cm
\epsfysize=6.0cm
\epsffile{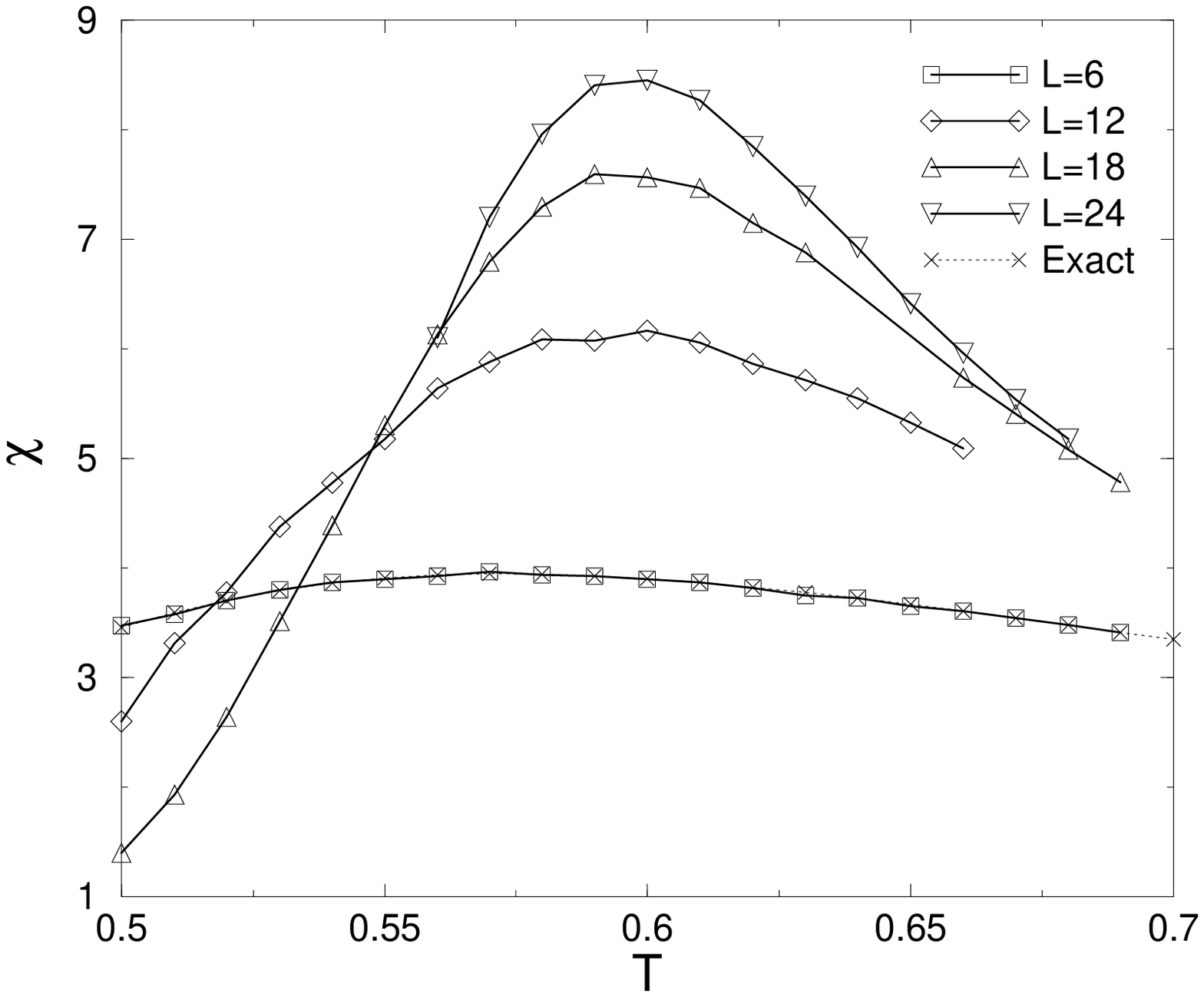}
\begin{figure}
\caption{\label{h2}  Staggered susceptibility $\chi$ versus
temperature $T$ for $h=0.25$, $J=1$. The data for system sizes
$L=6$, $12$, $18$ and $24$ were obtained from MC simulations using 
string dynamics. Exact transfer-matrix results for
$L=6$ are also shown. 
}  
\end{figure}}
\vbox{
\vspace{0.0cm}
\epsfxsize=8.0cm
\epsfysize=6.0cm
\epsffile{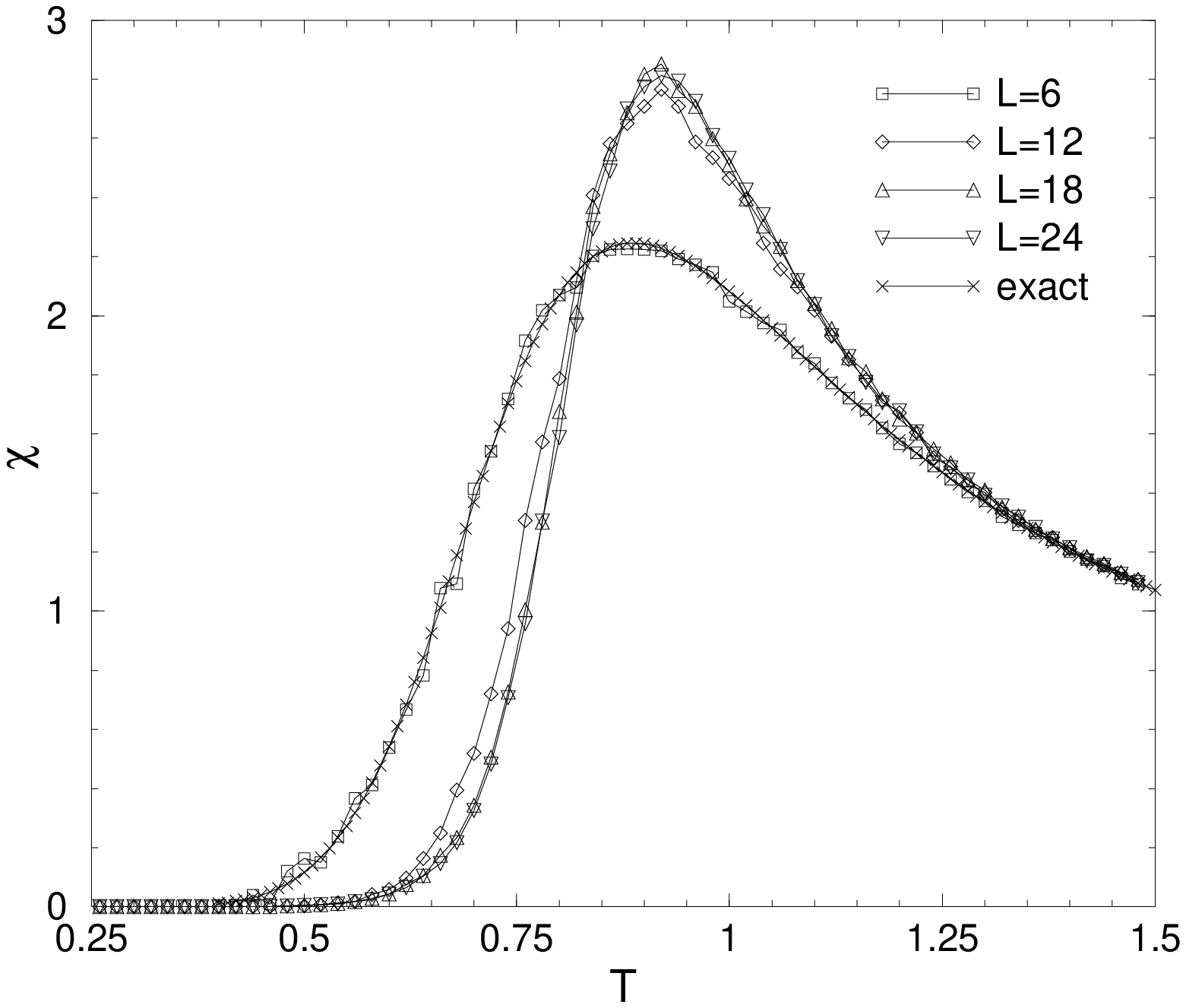}
\begin{figure}
\caption{\label{h3}  Staggered susceptibility $\chi$ versus
temperature $T$ for $h=0.4$, $J=1$. The data for system sizes
$L=6$, $12$, $18$ and $24$ were obtained from MC simulations using 
string dynamics. Exact transfer-matrix results for
$L=6$ are also shown.
}  
\end{figure}}
\vbox{
\vspace{0.5cm}
\epsfxsize=8.0cm
\epsfysize=6.0cm
\epsffile{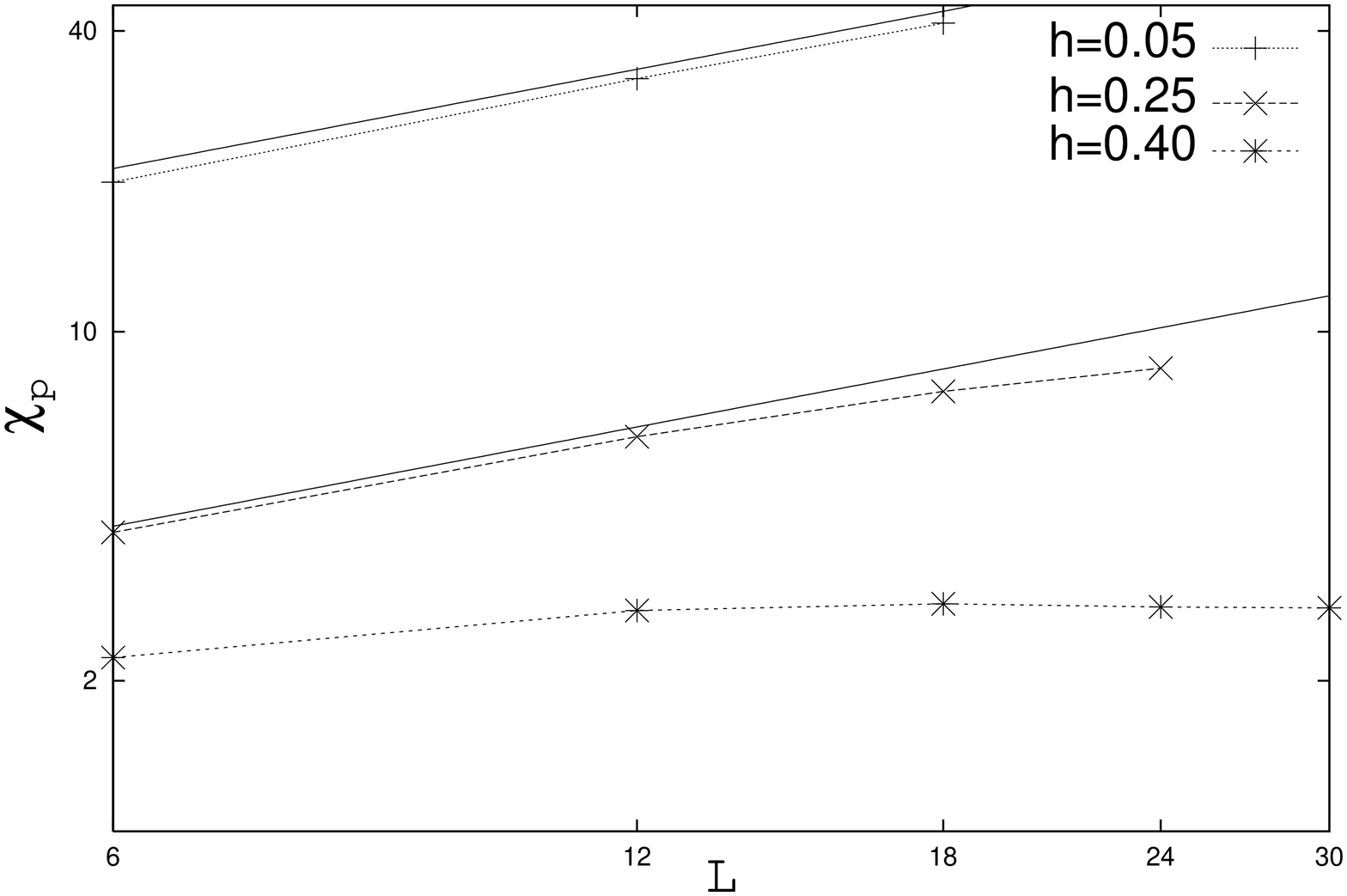}
\begin{figure}
\caption{\label{max} The susceptibility maximum $\chi_p$ plotted against
the system size $L$ for three different values ($0.05$, $0.25$ and
$0.4$) of the staggered field $h$. The solid lines correspond to the
power-law form $\chi_p \propto L^{2/3}$.}
\end{figure}}
peak of $\chi$ occurs
near $T \simeq 2.5h$, and the peak value of $\chi$ increases as $L$ is
increased. 
Finally, in Fig.~\ref{h3}, we have
shown the results for a high field value, $h=0.4$. In 
this case, equilibration times are quite small and we 
can simulate relatively large systems without any difficulty. 
All the MC data shown in Fig.~\ref{h3} were obtained with
averaging over only $2 \times 10^5$ MCS. 
We find that in this case, the staggered susceptibility saturates 
for $L \ge 12$, and clearly there is no phase transition. 

In Fig.~\ref{max}, we have plotted 
$\chi_p$, the value of the staggered susceptibility at the peak,
against the system size $L$ for the three different fields. 
As noted above, we get $\chi_p \sim L^{2/3}$ for $h=0.05$. For
$h=0.25$, the values of $\chi_p$ for $L=6$ and $L=12$ are consistent
with this power-law form, but the data for higher values of $L$ show
deviations from this form and signs of saturation. Finally, for
$h=0.4$, the peak value of $\chi$ clearly saturates for $L \ge 12$.

Taken at face value, these results would imply that for $J=1$, there is
a K-type transition for $h=0.05$, but no transition for $h=0.25$ and
$h=0.4$. In other words, there is a phase transition for small $h$,
which disappears beyond a critical value of the field. This naive
interpretation of the data is questionable because a line of continuous
phase transitions in the h-T plane is very unlikely to end abruptly
at some point
. A more plausible interpretation is that the system with
finite $J$ does not exhibit a true phase transition for any value of
the staggered field -- the signature of a phase transition found in the
scaling behavior of the data for small $h$ is a remnant of the
transition in the $J \to \infty$ limit. The behavior of a system with
finite $J$ would differ from that in the $J \to \infty$ limit only if
the values of the parameters $J$, $T$ and $L$ are such that excitations
out of the manifold ${\cal G}$ are not strongly suppressed. Since the
typical value of the local field in a configuration in $\cal G$ is
$2J$, the typical energy cost associated with a single-spin-flip excitation
out of this manifold is $4J$. Since this excitation can occur at any
site of the lattice, the free energy cost of such an excitation is
approximately given by $\delta F \simeq 4J-2T \ln L$. Such excitations
are likely to occur if $\delta F \le 0$, which corresponds to $L \ge
L_c = e^{2J/T}$. The values of $L_c$ at temperatures near the peak of
$\chi$ are $\approx 10^7, 28$ and $7.4$ for $h=0.05, 0.25$ and $0.4$,
respectively. In view of the very large value of $L_c$ for $h=0.05$, it
is not surprising that the MC results for $m$ and $\chi$ for $h=0.05$,
$J=1.0$, and $L \le 18$ are essentially identical to the results for
the same value of $h$ in the $J \to \infty$ limit. The power-law
scaling of the data for $\chi_p$ at $h=0.05$ can then be attributed to
the occurrence of a phase transition in the $J \to \infty$ limit. The
observation that for $h=0.25$, the numerical data for $\chi_p$ show 
deviations from power-law scaling with $L$ and signs of saturation 
for $L \ge 24$ is also consistent with this interpretation. The small 
value of $L_c$ for
$h=0.4$ implies that the effects of $J$ being finite should 
be evident even in the small samples we consider. The fact
that the data for $h=0.4$ clearly indicate the absence of any phase 
transition is, thus, consistent with the interpretation that there is
no phase transition for finite $J$.

While the scenario described above is consistent with all our numerical
data, we can not be absolutely sure that it is correct -- data for much
larger systems would be needed for a conclusive answer to the question
of whether a phase transition occurs for finite $J$. We note that even
if our interpretation is correct, the behavior of finite samples with
finite $J$ would look very similar to that near a true phase transition
if $h/J$ is small. In such cases, the value of $\chi_p$ will continue 
to grow with $L$ as a power law until $L$ becomes comparable to $L_c$, 
at which point $\chi_p$ will saturate. Since $L_c$ depends
exponentially on $J/h$, it would be very large for $h/J \ll 1$.

\subsection{ Cluster dynamics}
We have also performed simulations using a cluster method. We briefly
report our results here. This method
was introduced by Kandel {\it et al.}~\cite{kandel} for the study of 
frustrated systems. Recently Zhang and Cheng~\cite{zhang} have
applied this algorithm to the 
zero-field TIAFM. We have modified this algorithm to take into account
the presence of the staggered field. The cluster algorithm is usually
implemented in two steps. In the first step, one performs a ``freeze-delete''
operation on the bonds using a fixed set of rules~\cite{kandel,zhang}, 
which results in the
formation of independent clusters. The second step consists in flipping
these clusters. In our modified algorithm, the first step is unchanged.
The freeze-delete operations are exactly as in Ref.~\cite{zhang} and
are effected without considering the energy associated with the
staggered field. In the second step,
we calculate the staggered-field energy of every cluster and then flip it
using heat-bath rules. It can be proved that this procedure satisfies
the detailed balance condition.  
  
The cluster dynamics performs better than the single-spin-flip
dynamics and we have been able to obtain equilibrium averages for a
$L=6$ system ($J=1$, $h=0.05$) with $10^6$ MCS. However, for bigger
system sizes ($L\ge12$), we have not been able to achieve equilibration
even with runs over $10^8$ MCS. Thus this dynamics is much slower than
the string dynamics.
This is due to the following reason. While the cluster dynamics does 
allow the number of strings to change,  the 
clusters formed at low temperatures are quite large and the probability of 
flipping them becomes very small.     
In order to obtain quantitative comparisons of the three
different dynamics, we have studied the autocorrelation function, 
\bea
C(\tau)=\frac{ \langle M(\tau) M(0) \rangle - \langle M \rangle ^2}
{ \langle M^2 \rangle - \langle M \rangle ^2},
\eea
where $M$ is the total staggered magnetization and $\tau$ is the
``time'' measured in units of MCS.
In Figs.~\ref{time1} and \ref{time2}, we plot the results for $C(\tau)$ 
obtained from simulations using different dynamics at two different 
temperatures. The data correspond to a $L=6$ 
lattice and the averaging was carried out over $10^7$ 
MCS in all the cases. We note
that the single-spin-flip dynamics leads to a two-step relaxation -- a fast
one corresponding to equilibration within a sector and a slower one in
which different sectors are sampled.
The results shown in these figures also demonstrate the superiority of
the string dynamics over the other two methods at both high and low
temperatures. 
\vbox{
\vspace{0.5cm}
\epsfxsize=8.0cm
\epsfysize=6.5cm
\epsffile{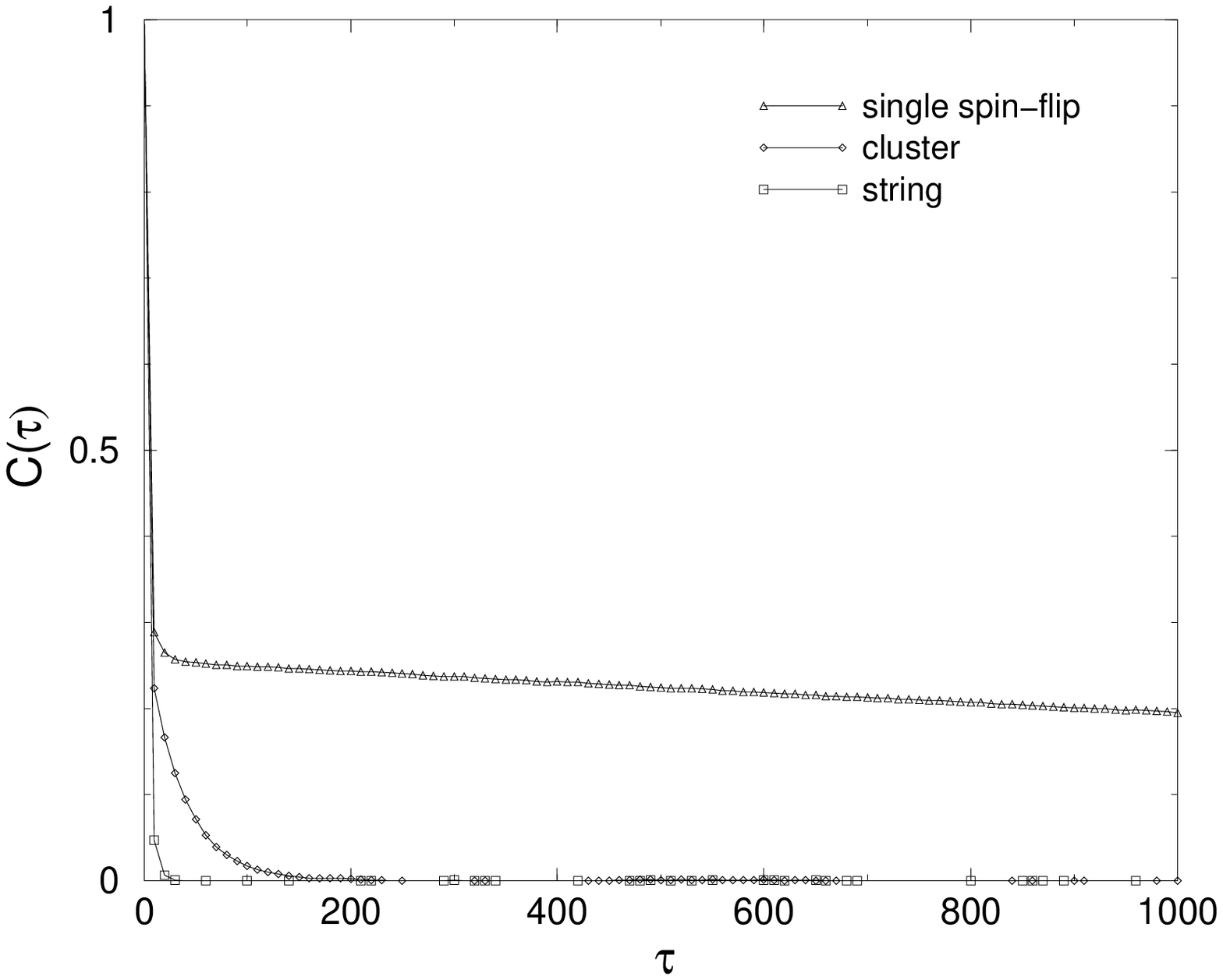}
\begin{figure}
\caption{\label{time1} Autocorrelation function $C(\tau)$ of the
staggered magnetization, obtained
from the three different dynamics at a comparatively high temperature,
$T=0.4$. The data are for a $6 \times 6$ sample with $J=1$, $h=0.05$.  
The ``time'' $\tau$ is measured in units of MC steps per spin.
}  
\end{figure}}
\vbox{
\vspace{0.5cm}
\epsfxsize=8.0cm
\epsfysize=6.5cm
\epsffile{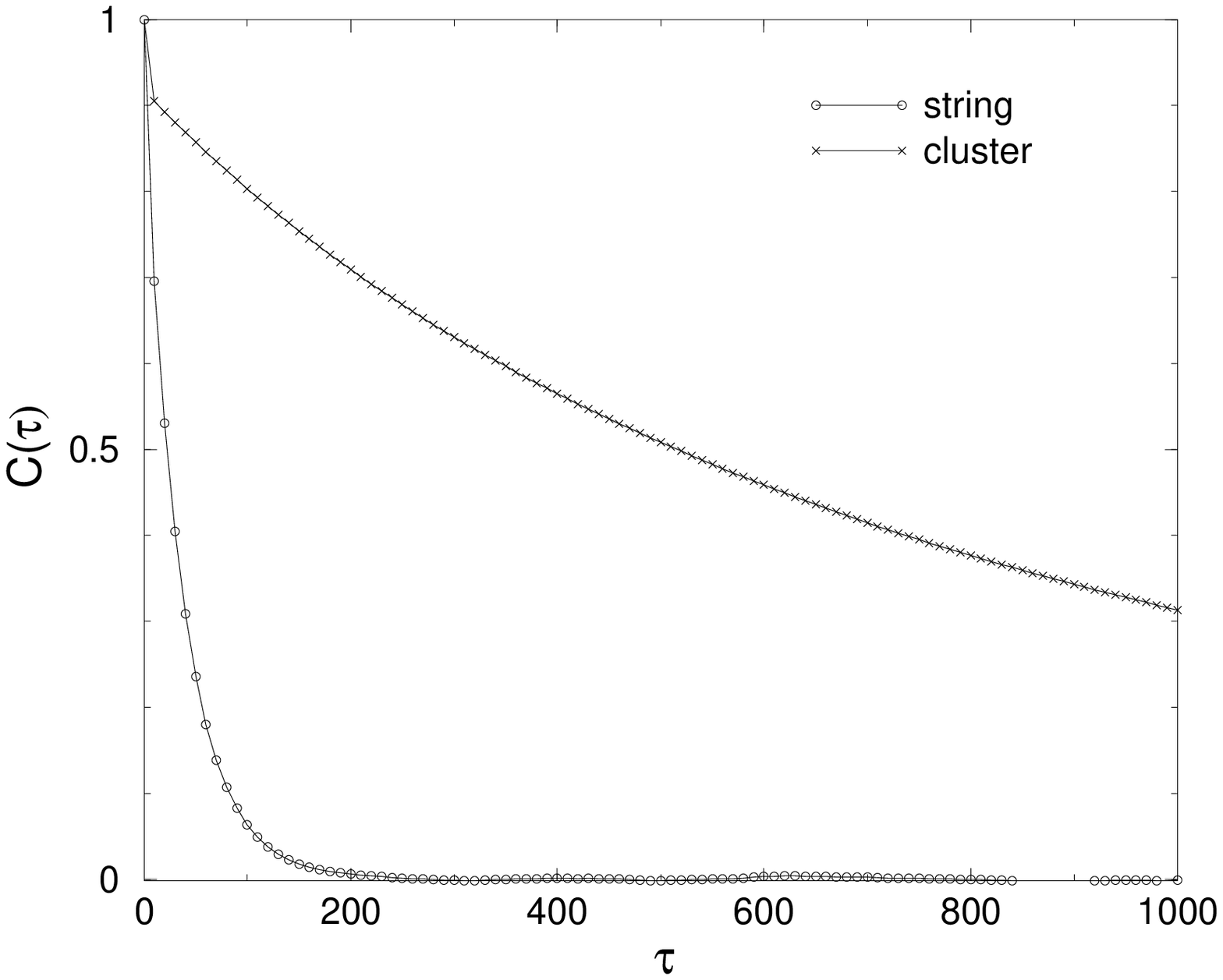}
\begin{figure}
\caption{\label{time2} Autocorrelation function $C(\tau)$ of the
staggered magnetization, obtained
from string and cluster dynamics at a low temperature,
$T=0.125$. The data are for a $6 \times 6$ sample with $J=1$, $h=0.05$.  
The ``time'' $\tau$ is measured in units of MC steps per spin.
}
\end{figure}}

\section{Summary and Discussion}

In summary, we have studied the equilibrium properties of a triangular
Ising antiferromagnet in the presence of an ordering field which is
conjugate to one of the degenerate ground states. We have addressed the
question of whether a phase transition can occur in this system.  Using
a mapping of the TIAFM ground states to dimer coverings, we find that
it is possible to obtain a very detailed description of the low-lying
energy states. In the limiting case of the coupling constant $J \to
\infty$, we show that the problem reduces to that of a  set of 
non-intersecting strings with long-range interactions. For this
limiting case, we prove existence of a transition which appears to be
$K$-type.  For finite $J$, we have studied the system using exact
numerical evaluation of the staggered magnetization and susceptibility
by transfer matrix methods, and also by MC simulations using three
different dynamics. We find that the dimer description also helps in
understanding the dynamics and in finding methods of improving the
efficiency of the MC simulation.  A single-spin-flip dynamics 
is very inefficient in
sampling different string sectors and at low temperatures, the system
stays stuck within a sector and shows thermodynamic behaviour
corresponding to that sector. A cluster dynamics method improves
over the single-spin-flip dynamics, but is still very slow at low
temperatures. We have developed a dynamics which allows moves that add
or remove pairs of strings. As expected, this greatly reduces
equilibration times. However, even with this increased efficiency, we
have not been able to equilibrate systems with $L>18$ in the
interesting region of low field values ($h/J << 1$). Hence our results
on possible phase transitions for finite $J$ are inconclusive, although
there are indications that a true phase transition does not occur for
finite $J$.

We close with a few comments on possible connections of the system
studied here with
supercooled liquids near the structural glass transition. The phase
transition we found in our model in the $J \to \infty$ limit is
similar in nature to the Gibbs-Di Marzio scenario~\cite{gdm} for the structural
glass transition. In the Gibbs-Di Marzio picture, the structural 
glass transition is supposed to be driven by 
an ``entropy crisis'' resulting from a vanishing of the configurational
entropy as the transition is approached from the 
high-temperature side. A similar vanishing of the entropy occurs at the
phase transition in our model. It is interesting to note in this
context that a ``compressible'' TIAFM model in which the ground-state 
degeneracy is lifted by a coupling of the spins with lattice degrees 
of freedom has been proposed~\cite{bulbul} as a simple spin model of 
glassy behavior. In view of these similarities with the structural 
glass problem, a detailed study of the dynamic behavior of our model 
would be very interesting. 

\section{Acknowledgements}
We thank Chinmay Das, Rahul Pandit and B. Sriram Shastry for 
helpful discussions.


\begin{references}
\bibitem[\dagger]{jnc} Also at the Condensed Matter Theory Unit,
Jawaharlal Nehru Centre for Advanced Scientific Research, Bangalore
560064, India.

\bibitem{wann}
G. H. Wannier, Phys Rev. {\bf 79}, 357 (1950).

\bibitem{stephen}
J. Stephenson, J. Math. Phys {\bf 5}, 1009 (1964).

\bibitem{nien}
B. Nienhuis, H. J. Hilhorst and H. W. J. Bl\"{o}te, J. Phys A:
Math. Gen. {\bf 17}, 3559 (1984). 

\bibitem{hout}
R. M. F. Houtappel, Physica {\bf 16}, 425 (1950).

\bibitem{kinz}
W. Kinzel and M. Schick, Phys. Rev. B {\bf 23}, 3435 (1981).

\bibitem{blote}
H. W. J. Bl\"{o}te and H. J. Hilhorst, J. Phys. A: Math. Gen. {\bf 15},
L631 (1982).

\bibitem{nighting}
H. W. J. Bl\"{o}te and M. P. Nightingale, Phys. Rev. B {\bf 47}, 15046
(1993).

\bibitem{silvio}
S. Franz and G. Parisi, Phys. Rev. Lett. {\bf 79}, 2486 (1997); M.
Cardenas, S. Franz and G. Parisi, J. Phys. A: Math. Gen. {\bf 31}, L183
(1998).

\bibitem{yang}
C. N. Yang and T. D. Lee, Phys. Rev. {\bf 87}, 410 (1952); M. Suzuki 
and M. E. Fisher, J. Math. Phys. {\bf 12}, 235 (1971).

\bibitem{mc} K. Binder and D. W. Herrman, {\it Monte Carlo Simulation
in Statistical Physics} (Springer-Verlag, Berlin, 1988).

\bibitem{fsc} J. L. Cardy, ed., {\it Finite Size Scaling} (Elsevier,
Amsterdam, 1988).

\bibitem{kandel}
D. Kandel, R. Ben-Av and E. Domany, Phys. Rev. B {\bf 45}, 4700
(1992). 

\bibitem{zhang}
G. M. Zhang and C. Z. Yang, Phys. Rev. B {\bf 50}, 12546 (1994). 

\bibitem{gdm} J. H. Gibbs and E. A. Di Marzio, J. Chem. Phys. {\bf 28},
373 (1958); G. Adams and J. H. Gibbs, J. Chem. Phys. {\bf 43}, 139
(1965). 

\bibitem{bulbul} L. Gu and B. Chakraborty, in {\it Proceedings of MRS
Symposium on Glasses}, eds. K. L. Ngai and C. A. Angell (1996)
[cond-mat 9612103].

\end{references}
\end{document}